\documentclass[useAMS,usenatbib,usegraphicx]{mn2e}

\usepackage{amssymb}
\usepackage{epsfig}
\usepackage{lscape}
\usepackage{graphicx}
\usepackage{rotating}
\usepackage{times}
\usepackage{color}
\bibpunct{(}{)}{;}{a}{}{,}

\newcommand{\rsi}{$\mathcal{R}({\rm Si})$}
\newcommand{\snia}{SN~Ia}
\newcommand{\sneia}{SNe~Ia}

\newcommand{\opar}{\hbox{$^{\scriptsize \rm  o}$}}
\newcommand{\oparsub}[1]{\hbox{$^{\scriptsize \rm  o}_{#1}$}}

\let\ts=\thinspace
\newcommand{\one}{\ts {\,\sc i}}
\newcommand{\two}{\ts {\,\sc ii}}
\newcommand{\three}{\ts {\,\sc iii}}

\newcommand{\nifs}{\ensuremath{^{56}\rm{Ni}}}
\newcommand{\cofs}{\ensuremath{^{56}\rm{Co}}}

\newcommand{\msun}{\ensuremath{\rm{M}_{\odot}}}
\newcommand{\lsun}{\ensuremath{\rm{L}_{\odot}}}
\newcommand{\kms}{\ensuremath{\rm{km\,s}^{-1}}}
\newcommand{\gcc}{\ensuremath{\rm{g\,cm}^{-3}}}

\def\cmfgen{{\sc cmfgen}}

\title[SN~Ia delayed-detonation models at maximum
  light]{One-dimensional delayed-detonation models of Type Ia
  supernovae: Confrontation to observations at bolometric maximum} 
\author[St\'ephane Blondin et al.]
{
St\'ephane Blondin,$^{1,2}$\thanks{E-mail: stephane.blondin@oamp.fr}
Luc Dessart,$^{1,3}$
D.~John Hillier,$^{4}$
and Alexei~M. Khokhlov$^{5}$\\
$^{1}$Aix Marseille Universit\'e, CNRS, LAM (Laboratoire d'Astrophysique
de Marseille) UMR 7326, 13388, Marseille, France.\\
$^{2}$Centre de Physique des Particules de Marseille (CPPM), Aix
Marseille Universit\'e, CNRS/IN2P3, Marseille, France.\\
$^{3}$ TAPIR, Mail code 350-17, California Institute of Technology,
Pasadena, CA 91125, USA.\\
$^{4}$Department of Physics and Astronomy, University
  of Pittsburgh, Pittsburgh, PA 15260, USA.\\
$^{5}$Department of Astronomy \& Astrophysics, the Enrico Fermi
Institute, and the Computation Institute, The University of Chicago,
Chicago, IL 60637, USA. 
}

\begin{document}

\date{Accepted 2012 November 23.  Received 2012 November 23; in original form 2012 October 15}

\pagerange{\pageref{firstpage}--\pageref{lastpage}} \pubyear{2012}

\maketitle

\label{firstpage}


\begin{abstract}
The delayed-detonation explosion mechanism applied to a
Chandrasekhar-mass white dwarf offers a very attractive model to
explain the inferred characteristics of Type Ia supernovae (\sneia).
The resulting ejecta are chemically stratified, have the same mass and
roughly the same asymptotic kinetic energy, but exhibit a range in
\nifs\ mass.  We investigate the contemporaneous photometric and
spectroscopic properties of a sequence of delayed-detonation models,
characterized by \nifs\ masses between 0.18 and 0.81\,\msun. Starting
at 1\,d after explosion, we perform the full non-LTE, time-dependent
radiative transfer with the code \cmfgen, with an accurate treatment
of line blanketing, and compare our results to \sneia\ at bolometric
maximum.  Despite the 1D treatment, our approach delivers an excellent
agreement to observations. We recover the range of
\snia\ luminosities, colours, and spectral characteristics from the
near-UV to 1\,$\mu$m, for standard as well as low-luminosity 91bg-like
\sneia.  Our models predict an increase in rise time to peak with
increasing \nifs\ mass, from $\sim$\,15 to $\sim$\,21\,d, yield peak
bolometric luminosities that match Arnett's rule to within 10\,\% and
reproduce the much smaller scatter in near-IR magnitudes compared to
the optical.  We reproduce the morphology of individual spectral
features, the stiff dependence of the \rsi\ spectroscopic ratio on
\nifs\ mass, and the onset of blanketing from Ti\two/Sc\two\ in
low-luminosity \sneia\ with a \nifs\ mass $\lesssim$\,0.3\,\msun.  We
find that ionization effects, which often dominate over abundance
variations, can produce high-velocity features in Ca\two\ lines, even
in 1D.  Distinguishing between different \snia\ explosion mechanisms
is a considerable challenge but the results presented here provide
additional support to the viability of the delayed-detonation model.
\end{abstract}



\section{Introduction}\label{sect:intro}

All models for Type Ia supernovae (\sneia) involve the thermonuclear
disruption of a C/O white dwarf (WD) star
\citep{Hoyle/Fowler:1960}. Fusion of $^{12}$C and $^{16}$O to
iron-peak elements provides the energy to unbind the star and
accelerate the ejecta to expansion velocities on the order of
10000\,\kms, while synthesizing $\lesssim1$\,\msun\ of \nifs\ to power
the subsequent light curve \citep{Colgate/McKee:1969}.

Nuclear combustion of C and O requires a large temperature
($\sim10^9$\,K) in the progenitor star. At present, these
conditions seem to be met when the WD mass approaches the
Chandrasekhar mass, $M_{\rm Ch}\approx1.4$\,\msun, through accretion
of material from a companion star (the so-called single-degenerate
scenario; \citealt{Whelan/Iben:1973}).  Alternatively, combustion may
be initiated when two C/O WDs coalesce (the double-degenerate
scenario; \citealt{Iben/Tutukov:1984,Webbink:1984}), in which case the
total mass of the system can differ significantly from $M_{\rm Ch}$.
While the single-degenerate scenario provides a natural explanation
for the apparent homogeneity of the \snia\ class, recent
simulations have shown that violent WD mergers are
compatible with both sub-luminous \citep{Pakmor/etal:2010} and normal
\citep{Pakmor/etal:2012} \sneia. It is thus not clear today whether
the Chandrasekhar mass represents a fundamental quantity for \sneia.

In addition to multiple progenitor channels, several explosion
mechanisms have been proposed. The propagation of the burning front
can proceed either sub-sonically (deflagration), or super-sonically
(detonation). Pure detonations of C/O WDs \citep[e.g.,][]{Arnett:1969}
burn the entire star to nuclear statistical equilibrium and fail to
synthesize the intermediate-mass elements (IMEs) at high velocities
needed to reproduce \snia\ spectra \citep[see,
  e.g.,][]{Branch/etal:1982}. Pure deflagrations
\citep[e.g.,][]{Nomoto/etal:1976} yield copious amounts of IMEs but
correspond to less energetic explosions, with \nifs\ masses typically
$\lesssim0.4$\,\msun \citep[see, e.g.,][]{Travaglio/etal:2004}, and
are thus considered a feasible mechanism only for the least luminous
events. More importantly, significant amounts of unburnt C/O are
expected to be mixed inwards to the WD centre \citep[see,
  e.g.,][]{Gamezo/etal:2003}, leading to the firm prediction of
conspicuous lines of O\one\ and C\one\ at nebular times that have
never been observed \citep{Kozma/etal:2005}.

To overcome the shortcomings of both explosion mechanisms,
\cite{Khokhlov:1991} proposed the delayed-detonation model in which
the explosion of a Chandrasekhar-mass C/O WD begins as a subsonic
deflagration near its centre and then switches to a supersonic
detonation wave at some density, $\rho_{\rm tr}$, via a
deflagration-to-detonation transition (or DDT). The initial
deflagration pre-expands the WD and creates low-density conditions
required for the production of intermediate mass elements. The ensuing
detonation incinerates the remaining unburnt matter.  By tuning
$\rho_{\rm tr}$ and the deflagration speed, one can
synthesize varying amounts of \nifs\ and hence account for the
observed variation in \snia\ peak luminosity, as well as reproduce the
correct stratification of chemical elements in the ejecta inferred
from spectroscopic observations. Additional variations in peak
luminosity can be induced by varying the C/O ratio and central density
of the progenitor WD star \citep{Hoeflich/etal:2010}.

The delayed-detonation model has been successfully tested against
\snia\ observations in its 1D
\citep[e.g.,][]{Hoeflich/Khokhlov/Wheeler:1995} and multi-D
\citep[e.g.,][]{KRW09} versions.  It also appears to be consistent
with observations of supernova remnants
\citep{Badenes/etal:2006,Badenes/etal:2008,Fesen/etal:2007}.  Using a
grid of 130 parameterized one-dimensional models,
\cite{Woosley/etal:2007} studied the ejecta properties necessary to
reproduce the fundamental properties of \snia\ light curves, and in
particular the width-luminosity relation
\citep{Pskovskii:1977,Phillips:1993}. They found a good agreement for
ejecta with a total burnt mass of $\sim$\,1.1\,\msun\ (leading to
comparable asymptotic kinetic energies in such models), provided the
inner layers were strongly mixed, and highlight the adequacy of the
delayed-detonation scenario in fulfilling these requirements.
However, despite a significant effort in three-dimensional simulations
\citep[e.g.,][]{Gamezo/Khokhlov/Oran:2005,Roepke/Niemeyer:2007}, it is
still unclear what may cause strong variations in transition density
in the DDT mechanism. Moreover, the lack of unburnt carbon in
delayed-detonation models is in conflict with observations of
C\two\,6580\,\AA\ absorption in $\sim30$\% of \sneia\ with
pre-maximum spectra
\citep{Parrent/etal:2011,Thomas/etal:2011,Folatelli/etal:2012,Silverman/Filippenko:2012}.

Differentiating between these various models requires detailed
radiative transfer simulations to match the observed evolution of
\snia\ light curves and spectra form early to late times. At present,
the observed properties of normal \sneia\ can be reproduced with
detonations in single sub-$M_{\rm Ch}$ WDs \citep{Sim/etal:2010} and
super-$M_{\rm Ch}$ double-WD mergers \citep{Roepke/etal:2012}, in
addition to the more conventional Chandrasekhar-mass
delayed-detonation model. This may mean that observational differences
between these models are too small, or that uncertainties in the
modeling approach are too large.

Radiative-transfer modeling of \snia\ ejecta is complicated by the
dominance of line opacity
\citep{Hoeflich/etal:1993,Pinto/Eastman:2000b}, with a mixed
absorptive/scattering character that is often enforced, particularly
in Monte-Carlo codes \citep[see,
  e.g.][]{Mazzali/Lucy:1993,SEDONA,Sim:2007}, rather than computed
directly as part of a non-Local-Thermodynamic-Equilibrium (non-LTE)
solution.  Other approaches treat some species in full non-LTE, while
others are treated in LTE
\citep[e.g.,][]{Hoeflich/etal:1998,Baron/etal:2012}.  Neglecting
time-dependent effects in the radiation transport, which are known to
be important
\citep[see][]{Hoeflich/Khokhlov:1996,Pinto/Eastman:2000b,DH10,Jack/etal:2009,Hillier/Dessart:2012},
numerous studies at bolometric maximum (and sometimes even later)
impose a diffusive inner boundary, even though some spectral regions
are no longer optically thick. This causes a systematic flux excess in
the red, i.e., photons erroneously injected at the inner diffusive
boundary fail to thermalize and directly escape the ejecta
\citep[e.g.,][]{Mazzali/etal:2008}.  The low 
densities and strong radiation field in \sneia\ is expected to drive
level populations out of LTE, indicating the need for a full non-LTE
treatment (\citealt{Baron/etal:1996}; Dessart et al., in prep.).

As part of a new and independent effort, we apply the \cmfgen\ code
\citep{HM98,Hillier/Dessart:2012} to the Chandrasekhar-mass
delayed-detonation model of \sneia. \cmfgen\ is a 1D, non-LTE,
time-dependent, radiative-transfer tool that allows for explicit
treatment of non-thermal processes and non-local energy deposition
(see \citealt{Hillier/Dessart:2012} for a full description of the code
in the context of SN calculations). In this paper we focus on the
photometric and spectroscopic properties of delayed-detonation models
at bolometric maximum, to study how well this explosion mechanism can
predict the observed diversity of \sneia\ at that well-defined time.

In section~\ref{sect:models} we present our grid of delayed-detonation
models and radiative transfer modeling. In Sect.~\ref{sect:strategy}
we present our overall strategy for comparing synthetic light curves
and spectra. We then discuss the photometric
(Sect.~\ref{sect:photprop}) and spectroscopic
(Sect.~\ref{sect:specprop}) properties at bolometric maximum. We
discuss comparisons to individual \snia\ observations in
Sect.~\ref{sect:spec_comp}. Discussion and conclusions follow in
Sect.~\ref{sect:ccl}.


\section{Numerical setup}\label{sect:models}


\subsection{Delayed-detonation models}\label{sect:ddt}

\begin{table*}
\caption{Delayed-detonation model parameters and nucleosynthetic yields for selected species.}\label{tab:modinfo}
\begin{tabular}{lc@{\hspace{1.8mm}}c@{\hspace{1.8mm}}c@{\hspace{1.8mm}}c@{\hspace{1.8mm}}c@{\hspace{1.8mm}}c@{\hspace{1.8mm}}c@{\hspace{1.8mm}}c@{\hspace{1.8mm}}c@{\hspace{1.8mm}}c@{\hspace{1.8mm}}c@{\hspace{1.8mm}}c@{\hspace{1.8mm}}c@{\hspace{1.8mm}}c@{\hspace{1.8mm}}c@{\hspace{1.8mm}}}
\hline\hline
\multicolumn{1}{c}{Model} & $\rho_{\rm tr}$ & $E_{\rm kin}$ & $v(\nifs)$  & $M$(\nifs) & $M$(Ni) & $M$(Co) & $M$(Fe) & $M$(Ti) & $M$(Ca) & $M$(S) & $M$(Si) & $M$(Mg) & $M$(Na) & $M$(O) & $M$(C) \\
 & [\gcc] & [B] & [\kms]  & [M$_{\sun}$] & [M$_{\sun}$] & [M$_{\sun}$] & [M$_{\sun}$] & [M$_{\sun}$] & [M$_{\sun}$] & [M$_{\sun}$] & [M$_{\sun}$] & [M$_{\sun}$] & [M$_{\sun}$] & [M$_{\sun}$] & [M$_{\sun}$] \\
\hline
DDC0           &   3.5(7) & 1.564 &  1.31(4) &      0.805 &      0.857 &   1.15(-2) &      0.160 &   1.19(-4) &   2.62(-2) &      0.101 &      0.160 &   3.51(-3) &   5.88(-6) &   5.29(-2) &   1.18(-3) \\
DDC6           &   2.7(7) & 1.503 &  1.21(4) &      0.727 &      0.768 &   9.10(-3) &      0.142 &   1.21(-4) &   3.06(-2) &      0.125 &      0.201 &   7.18(-3) &   9.66(-6) &   8.42(-2) &   1.76(-3) \\
DDC10          &   2.3(7) & 1.485 &  1.16(4) &      0.650 &      0.689 &   8.36(-3) &      0.137 &   1.24(-4) &   3.68(-2) &      0.149 &      0.234 &   9.55(-3) &   1.20(-5) &   9.67(-2) &   2.11(-3) \\
DDC15          &   1.8(7) & 1.468 &  1.12(4) &      0.558 &      0.596 &   7.87(-3) &      0.135 &   1.25(-4) &   4.17(-2) &      0.177 &      0.277 &   1.14(-2) &   1.63(-5) &      0.107 &   2.69(-3) \\
DDC17          &   1.6(7) & 1.410 &  1.09(4) &      0.458 &      0.495 &   7.24(-3) &      0.131 &   1.25(-4) &   4.45(-2) &      0.200 &      0.314 &   1.65(-2) &   2.43(-5) &      0.139 &   3.77(-3) \\
DDC20          &   1.3(7) & 1.382 &  1.04(4) &      0.344 &      0.380 &   6.43(-3) &      0.126 &   1.25(-4) &   4.53(-2) &      0.233 &      0.374 &   1.95(-2) &   3.40(-5) &      0.157 &   5.05(-3) \\
DDC22          &   1.1(7) & 1.301 &  9.85(3) &      0.236 &      0.271 &   5.52(-3) &      0.120 &   1.24(-4) &   4.04(-2) &      0.255 &      0.432 &   2.45(-2) &   6.07(-5) &      0.186 &   8.14(-3) \\
DDC25          &   8.0(6) & 1.177 &  8.36(3) &      0.180 &      0.215 &   4.37(-3) &      0.111 &   1.25(-4) &   2.34(-2) &      0.230 &      0.458 &   3.33(-2) &   1.46(-4) &      0.248 &   2.07(-2) \\
\hline
\end{tabular}

\flushleft
{\bf Notes:}
Numbers in parenthesis correspond to powers of ten.
The ratio of the deflagration velocity to the local sound speed ahead of the flame is $\alpha=0.03$ for all models;
$\rho_{\rm tr}$ is the transition density at which the deflagration is artificially turned into a detonation;
$E_{\rm kin}$ is the asymptotic kinetic energy (units: $1\mathrm{B} \equiv 1 \mathrm{Bethe} = 10^{51}$\,erg);
$v(\nifs)$ is the velocity of the ejecta shell that bounds 99\% of the total \nifs\ mass (see Sect.~\ref{sect:ddt} for a discussion).
\end{table*}

\begin{figure}
\centering
\epsfig{file=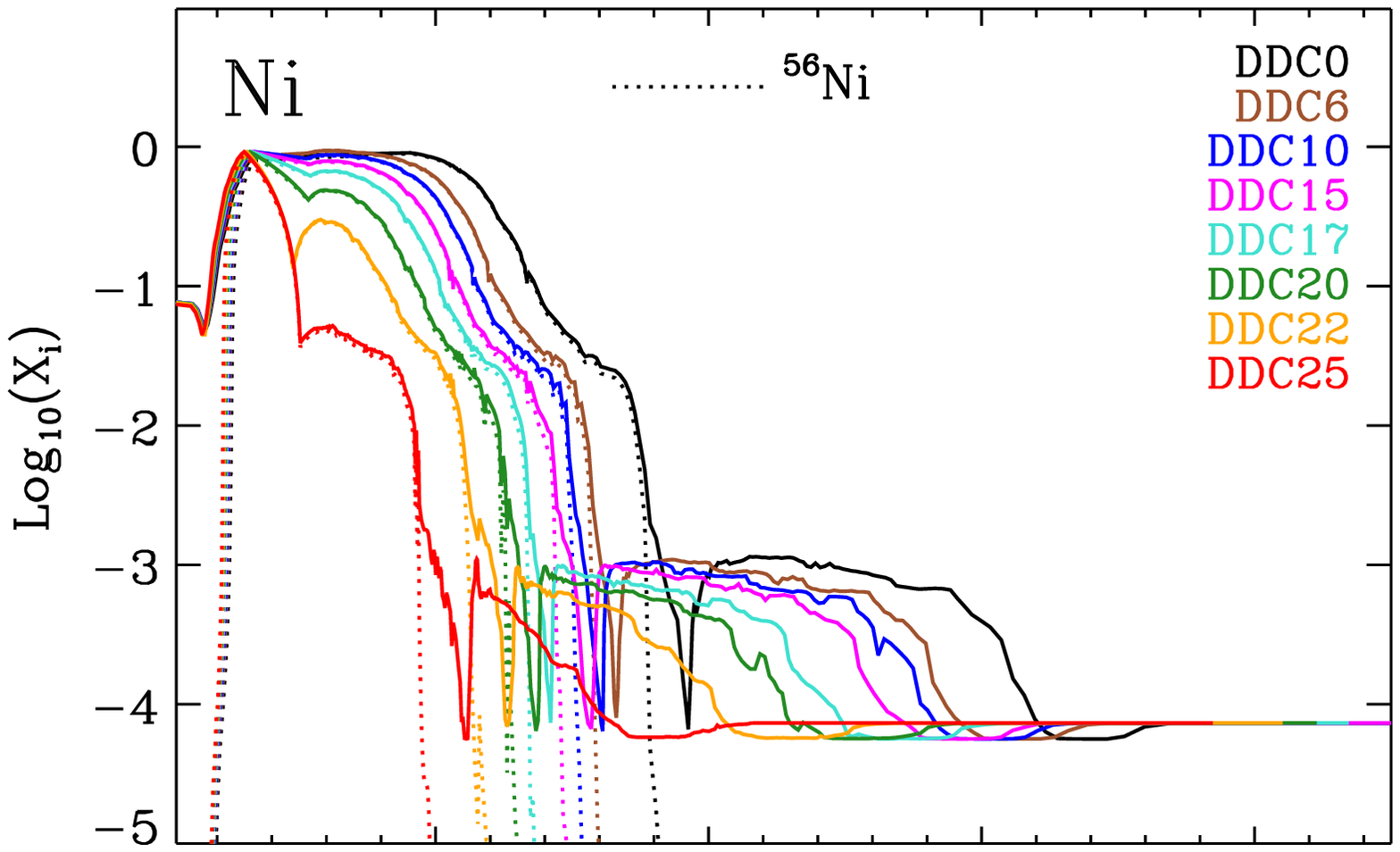,width=8.45cm} 
\epsfig{file=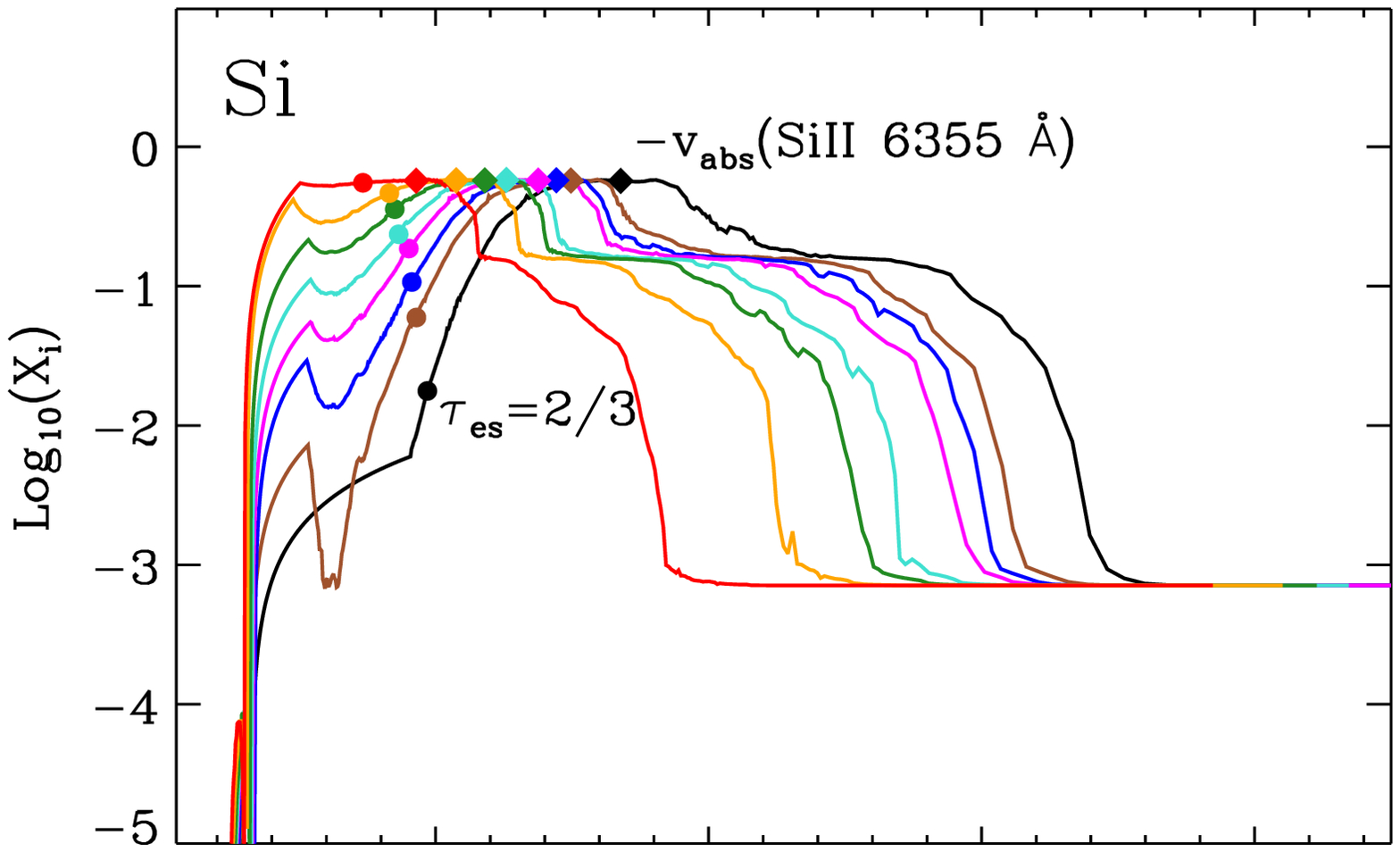,width=8.45cm} 
\epsfig{file=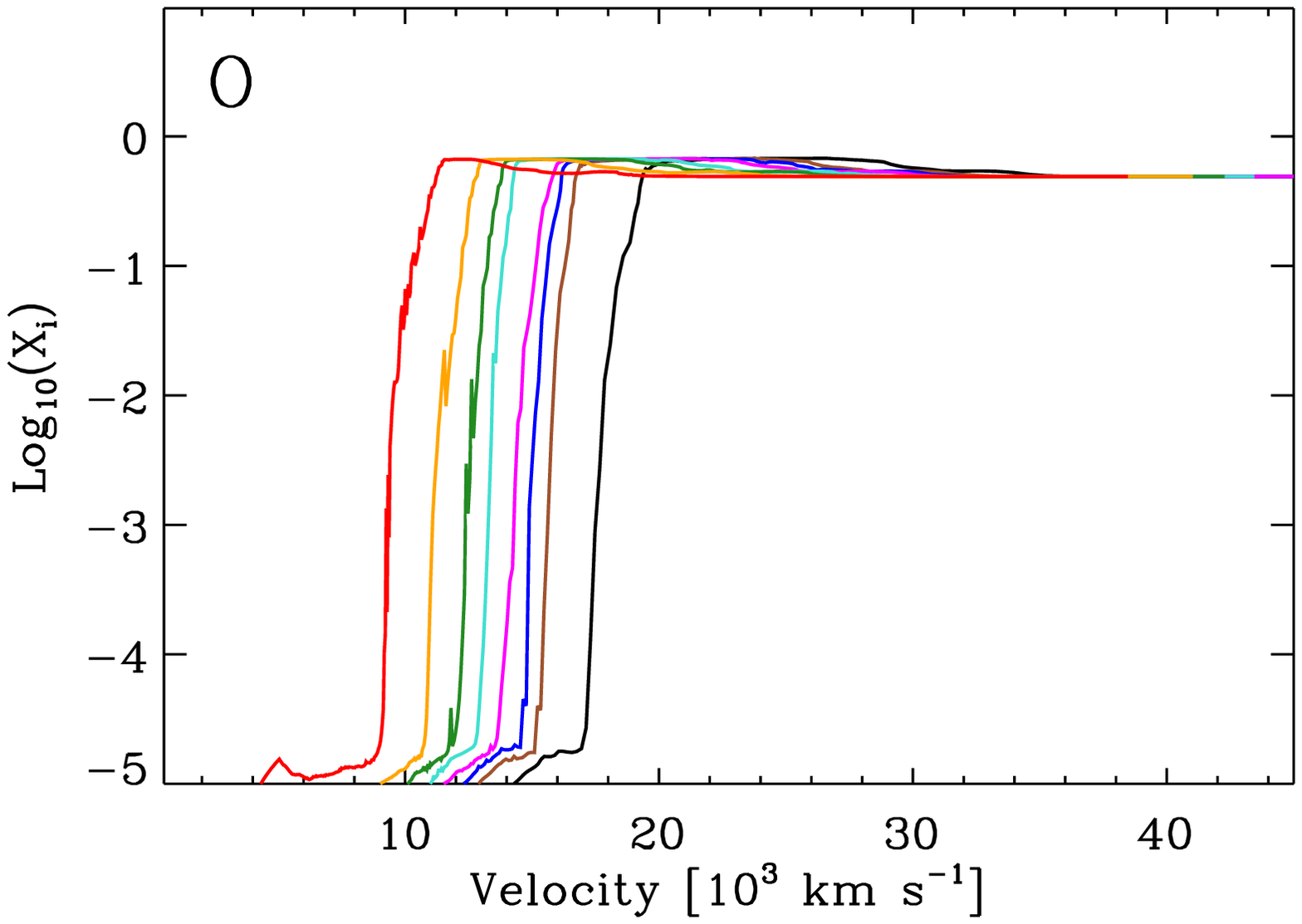,width=8.45cm}
\caption{\label{fig:comp_elem_distrib}
Comparison of the mass fractions of Ni (top; \nifs\ shown as dotted
lines), Si (middle), and O (bottom) in velocity space for our model
set. The drop in \nifs\ mass fraction (e.g., at
$\sim$\,5000\,\kms\ in model DDC25) corresponds to the ejecta location
where the deflagration transitions to a detonation. Note the absence
of \nifs\ in the inner ejecta layers (up to $v\sim$\,2000\,\kms) due
to rapid electron captures. 
In the middle panel we overplot  the
location where the inward-integrated electron scattering optical depth
$\tau_{\rm es}=2/3$ (dots), as well as minus the velocity at maximum
absorption in the Si\two\,6355\,\AA\ line (negative by convention;
diamonds).
}
\end{figure}

A grid of one-dimensional delayed-detonation models was calculated
following the method described in \cite{Khokhlov:1991}.  
The starting point for all models is a Chandrasekhar-mass
WD star (1.41\,\msun) with a central density of $2.6\times10^9$\,\gcc,
composed of equal mass fractions of $^{12}$C and $^{16}$O
($X\approx0.5$) with traces of $^{22}$Ne ($X\approx1.4\times10^{-2}$)
and solar abundances for all other elements \citep{Anders/Grevesse:1989}.
The explosion hydrodynamics is described by the reactive flow Euler
equations of fluid dynamics which are solved with a one-dimensional
Lagrangian hydrodynamics code. During the explosion, nuclear energy
release and neutronization of matter were described using a
five-equation reaction scheme for $^{12}$C concentration, electron and
mean ion mole fractions, nuclear binding energy, and the nuclear
statistic equilibrium (NSE) progress variable. The reaction scheme takes into
account the energy release during the major stages of explosive
carbon-oxygen burning: (1) burning of carbon to oxygen, neon,
magnesium and silicon, (2) burning of oxygen and establishment of
nuclear quasi-equilibrium in Si-group nuclei, and finally (3) the onset of
NSE during which the Si-group elements are transformed into iron-peak
nuclei. The scheme takes into account neutronization and neutrino
energy losses due to electron captures on iron-group nuclei. 
The neutronization rates were updated according to
\cite{Seitenzahl/etal:2009}.

The equation of state includes contributions from electrons and
positrons with an arbitrary degree of degeneracy and relativism, ideal
ions, and equilibrium radiation. Screening of the $^{12}$C$+^{12}$C
nuclear reaction is taken into account but Coulomb corrections to the
equation of state are ignored
\citep[see][]{Bravo/Garcia-Senz:1999}. They change the relation
between the initial central density and the WD mass by
$\sim$2--3\% but do not noticeably affect the explosion predictions.
The approximate kinetic scheme neglects differences between the
quasi-equilibrium energy release in Si-group elements and the NSE
energy release in Fe-peak elements and may overestimate the energy
generation rate during oxygen burning to silicon by up to
$\sim$20\% as compared to detailed nuclear kinetics.

The initial WD is assumed to be in hydrostatic equilibrium, and the
discretized equations of energy and nuclear kinetics are solved
simultaneously with high precision to prevent excitation of artificial
pulsations during the WD explosion.  Weak interaction rates and
neutrino energy losses are taken from \cite{Seitenzahl/etal:2009}.
Detailed nucleosynthesis is calculated by post processing temperature,
density, and neutronization histories of fluid elements with a
detailed reaction network consisting of 144 individual nuclei, ranging
from protons and neutrons to zinc. Forward reaction rates, partition
functions, and binding energies were taken from
\cite{Fowler/etal:1978}, \cite{Woosley/etal:1978}, and F.~Thielemann
(1993, private communication). Reverse reactions were calculated from
the principle of detailed balance.

The deflagration is propagated numerically with a prescribed velocity
$D_{\rm def}=\alpha a_s$ relative to unburnt matter, where $\alpha < 1$
is a constant and $a_s$ is the local sound speed ahead of the
deflagration front.  We use $\alpha=0.03$ in the models considered in
this paper.  The transition to a detonation is artificially triggered by
rising $D_{\rm def}$ above the maximum Chapman-Jouguet speed for a
steady-state deflagration when the density ahead of the deflagration
front reaches a prescribed transition density, $\rho_{tr}$. This
causes the onset of a self-sustained supersonic detonation wave after
the accelerated deflagration front propagates $\sim$\,3--5
computational cells.  The calculations are run until 30--60\,s past
explosion, by which time the ejecta have expanded by two orders of
magnitude and are in homologous expansion.

The earlier the transition from a deflagration to a detonation
occurs (i.e. the higher the transition density $\rho_{\rm tr}$), the
more efficiently the burning proceeds to large velocities with more
\nifs\ being synthesized at the expense of intermediate-mass
elements (Fig.~\ref{fig:comp_elem_distrib}). From model DDC0 to
DDC25, the velocity of the ejecta shell that bounds 99\% of the total
\nifs\ mass drops from $\sim$\,13000\,\kms\ to $\sim8000$\,\kms, while
the mass of synthesized \nifs\ goes from $\sim$\,0.81\msun\ to
$\sim$\,0.18\,\msun\ (see Table~\ref{tab:modinfo}).
At the largest velocities one recovers the
original composition of the progenitor WD star.

Since the WD is almost entirely consumed in 1D delayed-detonation
models (there is at most $\sim0.02$\,\msun\ of unburnt carbon for
model DDC25), the fraction of primordial C/O not burnt to \nifs\ is
burnt to IMEs, with a comparable nuclear energy release in both
cases. This leads to a unique property of the delayed-detonation model, 
whereby variations in the transition density results in
\snia\ explosions with the same ejecta mass and roughly the same
asymptotic kinetic energy ($E_{\rm kin}$), but with varying amounts of
\nifs. In our model set, $E_{\rm kin}$ varies by only 50\% for a
change in \nifs\ mass by a factor of 4.5.


\subsection{Non-LTE time-dependent radiative transfer}\label{sect:rt}

Our approach to solve the non-LTE time-dependent radiative-transfer
problem for SN ejecta is versatile and generic. It applies
irrespective of the SN type and we thus use here the same approach as,
e.g., in the recent study of pair-instability SN explosions by
\cite{Dessart/etal:2012e}.  The technique is presented in a sequence
of papers
\citep{DH05a,Dessart/Hillier:2008,Li/etal:2012,Dessart/etal:2012a}.
A thorough discussion of the application of the code \cmfgen\ to
supernovae has been presented by \citet{Hillier/Dessart:2012}.

Starting at 1\,d after explosion, we simultaneously simulate the
evolution of the full ejecta, including the gas thermodynamics state
(non-LTE ionization and excitation, and temperature) and the radiation
(either trapped or escaping).  This evolution is controlled by cooling
and heating processes. At early times, cooling stems primarily
from expansion, which in \sneia\ corresponds to a factor of 10$^{6-7}$
increase in radius from the original exploding white dwarf to the time
of bolometric maximum at $\sim$\,20\,d.  Heating is dominated in our
simulations by the radioactive decay of \nifs\ and \cofs\ nuclei. 
    
We solve the radiative-transfer problem using a non-LTE
treatment. Hence, all atom/ion level populations are determined
through a solution of the statistical equilibrium equations, coupled
to the gas energy equations and the 0$^{\rm th}$ and 1$^{\rm st}$
moments of the radiative-transfer equation.  Level populations are
therefore controlled by radiative and collisional rates explicitly
(see \citealt{HM98} for details). We do not use the questionable
assumption of LTE for the thermodynamical state of the gas, nor the
Equivalent-Two-Level-Atom approximation for solving level populations.
Together with charge conservation, this yields a non-LTE ionization.
Furthermore, we allow for time dependence both for the radiative
transfer, in order to compute accurately the transport of radiation
through the ejecta, as well as in the statistical-equilibrium and
energy equations, in order to account for the time-dependent effects
on the non-LTE ionization/excitation state of the gas
(\citealt{Dessart/Hillier:2008}; see also
\citealt{Utrobin/Chugai:2005}).

Opacity plays a critical role in SNe Ia. Through its impact on the
ejecta optical depth, it controls in part the light-curve morphology
\citep{Arnett:1982a,Hoeflich/etal:1993,Pinto/Eastman:2000b}.  In the
spectrum formation region, it modulates the magnitude of line
blanketing \citep{DH10,Dessart/etal:2012e}, and is thought in SNe Ia
to lead to a secondary maximum in near-IR light curves
\citep{Kasen:2006}.  In \cmfgen, we incorporate line blanketing
by explicitly treating bound-bound transitions between model
atom/ion levels (this is part of the non-LTE solution).  Hence, we
use no ad-hoc prescription for the thermalization character of the
gas, as typically done in the community (sometimes in combination with
LTE for the state of the gas). In the present simulations, we adopt a
set of model atoms for which we obtain converged results (i.e.,
increasing further the size of the model atoms alters negligibly the
resulting radiation and gas properties) --- this required extensive
testing, whose results will be presented in a separate paper (Dessart
et al., in prep).  In practice, we include C\one--{\sc iv},
O\one--{\sc iv}, Ne\one--{\sc iii}, Na\one, Mg\two--{\sc iii},
Al\two--{\sc iii}, Si\two--{\sc iv}, S\two--{\sc iv}, Ar\one--{\sc
  iii}, Ca\two--{\sc iv}, Sc\two--{\sc iii}, Ti\two--{\sc iii},
Cr\two--{\sc iv}, Mn\two--{\sc iii}, Fe\one--{\sc vii}, Co\two--{\sc
  vii}, and Ni\two--{\sc vii}.  The number of levels (both
super-levels and full levels; see \citealt{HM98} for details) is given
in the appendix, in Table~A1.  The highest ionization
stages that are needed to model the transport at early times are no
longer needed at bolometric maximum and thus these ions are omitted
from that table.

The output of the hydrodynamical calculations at 30--60\,s past
explosion is evolved to 1\,d using a separate program.
This program solves the energy equation given by the first law of
thermodynamics, assuming the material is radiation dominated (which
holds everywhere apart from the \nifs-deficient higher-density inner
ejecta below $\sim$\,2000\,\kms); we allow for \nifs\ decay but
neglect the diffusion of the associated heat. Given the high ejecta
densities during the first day after explosion, and our focus on
maximum-light spectra $\sim$\,20\,d later, these approximations are
adequate.

We then remap the resulting ejecta at 1\,d into \cmfgen. The ejecta,
whose velocity ranges from 500 to 45000\,\kms\ and density from
$\sim$\,3$\times$\,10$^{-9}$ to $\sim$\,3$\times$\,10$^{-15}$\,\gcc, 
are typically resolved with 100-120 radial points, distributed evenly
in optical-depth scale to have a minimum of 5--7 points per decade. 
For the initial relaxation from the hydrodynamical input (see below),
we keep the temperature fixed and solve for the full relativistic
transfer in non-LTE \citep{Hillier/Dessart:2012}. The  inner ejecta is
very hot at that time (typically 10$^5$\,K in the innermost \nifs-rich
region), while the outer ejecta is at
$\lesssim$\,10$^3$\,K. Since we model the full ejecta, this requires
the simultaneous treatment of highly ionized species (which we limit
to six-times ionized Fe/Co/Ni and five-times ionized O, for example)
and neutral species with C\one, O\one, or Fe\one. To avoid numerical
overflow in this initial relaxation at 1\,d (see details in
\citealt{DH10}), we have to enforce a floor temperature in the outer
ejecta of 6000\,K --- this floor value is reduced to 2000--3000\,K in
subsequent time steps when we let the temperature vary in the
time-dependent calculation. Together with the other approximations
used for the evolution of the gas from $\lesssim$\,100\,s to 1\,d,
this implies that our simulations need a few time steps to relax from
their initial conditions.

We adopt a time increment of 10\% of the current time until
$\sim$\,21\,d past explosion, corresponding to 33 time steps from the
initial time at 1\,d. Subsequently, we use a fixed 2\,d increment to
ensure a good time resolution.  In the models with a large
\nifs\ mass, the outer ejecta relaxes to the outward diffusion of heat
released at greater depths, as well as radiative and expansion
cooling, within a few 0.1\,d because of the close proximity of the
\nifs\ to the outer ejecta. In the models with a low \nifs\ mass, this
relaxation is slower, and takes $\sim$\,1--2\,d.  None of these issues
are important for the comparison to observations at bolometric
maximum, $\sim$\,15--21\,d after explosion.  Probably more relevant
than any shortcoming of this initial setup is the neglect of
multi-dimensional effects, in particular associated with inward mixing
of unburnt C/O and outward mixing of \nifs\ \citep[see,
  e.g.,][]{Gamezo/Khokhlov/Oran:2005}.

We allow for non-local energy deposition from radioactive decay, using
a Monte-Carlo approach for $\gamma$-ray transport
\citep{Hillier/Dessart:2012}. We also treat the non-thermal processes
associated with the high-energy electrons produced by Compton
scattering and photoelectric absorption of these $\gamma$-rays
\citep{Li/etal:2012,Dessart/etal:2012a}.  These processes are
generally unimportant at early times and so we included them starting
at $\sim$\,15\,d after explosion.


\section{Strategy}
\label{sect:strategy}

In this study, we adopt a forward-modeling approach. We start with a
grid of realistic explosion models and solve the time-dependent
radiation-transport problem.  An important validation of such
delayed-detonation models is through direct confrontation to
observations. We do not attempt to model a specific SN --- instead we
confront each of our models with a large database of SN~Ia light
curves and spectra.  Our goal is to explore how well our range of
delayed-detonation models (which span a factor of 4.5 in \nifs\ mass)
can reproduce the observed range of SN~Ia properties, as for example
described by the different subclasses proposed by
\citet{Branch/etal:2006}.

All models were evolved from 1\,d after explosion until beyond
bolometric maximum. Here we focus the discussion exclusively on the
time of bolometric maximum, and defer to subsequent papers the study
of the rise to peak, the early post-peak evolution and the
width-luminosity relation, as well as the nebular phase.
The spectra at each time step are integrated over frequency to yield
bolometric light curves.  We infer the time of peak bolometric
luminosity through a 3$^{\rm rd}$-order polynomial fit to the light
curve. The synthetic spectra are interpolated in time to yield
maximum-light spectra for all models.  Since we simultaneously compute
spectra and light curves, we can directly compare
SN luminosity and spectroscopic morphology. For educational purposes,
we can reveal the direct contributions of individual ions to
the total flux by computing the synthetic spectra using the same
converged radiative-transfer solution, but ignoring all bound-bound
transitions of a given ion in the formal solution of the transfer
equation (see Appendix~\ref{sect:ladder}).

In the next sections, we present the basic properties of our
delayed-detonation model series and confront their photometric
(Sect.~\ref{sect:photprop}) and spectroscopic
(Sect.~\ref{sect:specprop}) properties to the plethora of
publicly-available \snia\ observations.  Since the true bolometric
luminosity is not known for most \sneia, we generate pseudo-bolometric
(UVOIR) light curves from $UBVRI$ magnitudes for both models and data
using the method of \cite{Valenti/etal:2008}. Observed magnitudes are
corrected for extinction (in the MW as well as in the host galaxy) and
placed on an absolute scale using the distance moduli inferred for
each SN (Sect.~\ref{sect:spec_comp}).  We then compare the synthetic
and observed peak UVOIR luminosities, the ratio of which is noted
$Q_{\rm uvoir}$. We also compare the magnitudes and broadband colours
at bolometric maximum in various optical and NIR bands.  Observed
spectra are de-redshifted and de-reddened and scaled to the absolute
$V$-band magnitude derived from photometry.  To enable a better
comparison of individual spectroscopic features, an additional
scaling, $F_{\rm scl}$, is applied to the data in order to match the
mean synthetic flux in the range 4000--6000\,\AA. In most cases
$F_{\rm scl}\approx 1$ to within 10\%, and in all cases consistent
with unity when taking into account errors on the adopted distances.
The detailed comparison to individual observations achieving the
``best-match" to each of our delayed-detonation models is presented in
Sect.~\ref{sect:spec_comp}.


\section{Photometric properties at bolometric maximum}\label{sect:photprop}

\begin{table*}
\caption{Various quantities at bolometric maximum characterizing our delayed-detonation model set.}\label{tab:ejectaprop}
\begin{tabular}{l@{\hspace{2.0mm}}c@{\hspace{2.0mm}}c@{\hspace{2.0mm}}c@{\hspace{2.0mm}}c@{\hspace{2.0mm}}r@{\hspace{2.0mm}}c@{\hspace{2.0mm}}c@{\hspace{2.0mm}}c@{\hspace{2.0mm}}c@{\hspace{2.0mm}}c@{\hspace{2.0mm}}c@{\hspace{2.0mm}}c@{\hspace{2.0mm}}c@{\hspace{2.0mm}}c@{\hspace{2.0mm}}c@{\hspace{2.0mm}}}
\hline\hline
\multicolumn{1}{c}{Model} & $M(\nifs)$ & $t_{\rm rise}$ & $L_{\rm bol}$   & $Q_\gamma$ & $B-R$       & $M_\tau$       & $v_\tau$        & $T_\tau$     & $X({\nifs})_0$ & $X$(Fe) & $X$(Ti) & $X$(Sc) & $X$(Ca) & $X$(Si) & $X$(O) \\
                          &[M$_{\sun}$]& [day]          & [erg\,s$^{-1}$]  &            & [mag]       & [\msun]        & [\kms]         & [$10^4$\,K]         &               &  &  &  &  &  &  \\
\hline
DDC0           & 0.805 & 15.71 & 1.77(43) & 0.91 &  $-$0.17 & 0.848 &  9699 & 1.527 & 8.46(-1) &  1.18(-1) &  1.58(-5) & 5.58(-11) &  2.52(-2) &  4.00(-2) &  2.76(-7) \\
DDC6           & 0.727 & 16.89 & 1.62(43) & 0.98 &  $-$0.13 & 0.809 &  9297 & 1.422 & 7.24(-1) &  1.15(-1) &  2.14(-5) & 8.63(-11) &  4.00(-2) &  8.95(-2) &  6.83(-7) \\
DDC10          & 0.650 & 17.44 & 1.45(43) & 1.01 &  $-$0.09 & 0.793 &  9129 & 1.342 & 6.13(-1) &  1.08(-1) &  2.38(-5) & 1.10(-10) &  4.89(-2) &  1.36(-1) &  1.12(-6) \\
DDC15          & 0.558 & 17.76 & 1.22(43) & 1.03 &  $-$0.02 & 0.786 &  9019 & 1.263 & 4.54(-1) &  9.27(-2) &  2.43(-5) & 1.48(-10) &  5.60(-2) &  2.14(-1) &  2.05(-6) \\
DDC17          & 0.458 & 18.90 & 1.03(43) & 1.10 &     0.05 & 0.738 &  8647 & 1.197 & 3.68(-1) &  8.69(-2) &  2.43(-5) & 1.60(-10) &  5.94(-2) &  2.56(-1) &  2.52(-6) \\
DDC20          & 0.344 & 19.38 & 7.65(42) & 1.11 &     0.15 & 0.726 &  8504 & 1.091 & 1.95(-1) &  6.70(-2) &  2.02(-5) & 2.10(-10) &  5.65(-2) &  3.62(-1) &  4.46(-6) \\
DDC22          & 0.236 & 19.52 & 5.10(42) & 1.09 &     0.65 & 0.721 &  8306 & 0.980 & 8.71(-2) &  5.09(-2) &  1.50(-5) & 2.69(-10) &  4.68(-2) &  4.61(-1) &  7.69(-6) \\
DDC25          & 0.180 & 20.50 & 3.64(42) & 1.06 &     0.96 & 0.626 &  7338 & 0.920 & 3.49(-2) &  4.06(-2) &  9.58(-6) & 3.44(-10) &  3.29(-2) &  5.51(-1) &  2.18(-5) \\
\hline
\end{tabular}

\flushleft
{\bf Notes:} Numbers in parentheses correspond to powers of ten. $Q_\gamma$ is the ratio of the peak bolometric luminosity to the instantaneous rate of decay energy. The Lagrangian mass coordinate $M_\tau$, velocity $v_\tau$, temperature $T_\tau$, and mass fractions $X$ are given at the location where the inward-integrated electron-scattering optical depth $\tau_{\rm es}=2/3$.
Note that the reported \nifs\ mass fraction corresponds to a time immediately after explosive burning ceases, rather than the time of bolometric maximum.
\end{table*}

The fundamental asset of the delayed-detonation model
is its ability to induce a variation in \nifs, controlled through the
deflagration-to-detonation transition density. This introduces
diversity in ejecta that have otherwise very similar credentials
(mass, kinetic energy, energy lost to unbind the progenitor, energy
lost to expansion).  It is thus no surprise that this variation in
\nifs\ drives the main contrast in model results.

The first observable affected by the variation in \nifs\ mass is the
peak luminosity, which correlates with the heating rate from decay.
In the sequence DDC25 to DDC0, the peak bolometric luminosity
ranges from 3.64$\times$10$^{42}$\,erg\,s$^{-1}$ to
1.77$\times$10$^{43}$\,erg\,s$^{-1}$ (i.e., from
9.5$\times$10$^{8}$\,\lsun\ to 4.6$\times$10$^{9}$\,\lsun).  We obtain
the following relation for our model set between the peak bolometric
luminosity and the \nifs\ mass:
$$ L_{\rm bol,peak} \sim \, 2.21 \times 10^{43} \times
\frac{M(\nifs)}{\msun} \,{\rm erg}\,{\rm s}^{-1} \, .
$$ 
The degeneracy of progenitor and explosion properties,
varying primarily in the amount of heating through
decay, is in part the cause of this linear relation to \nifs\ mass.

\begin{figure}
\epsfig{file=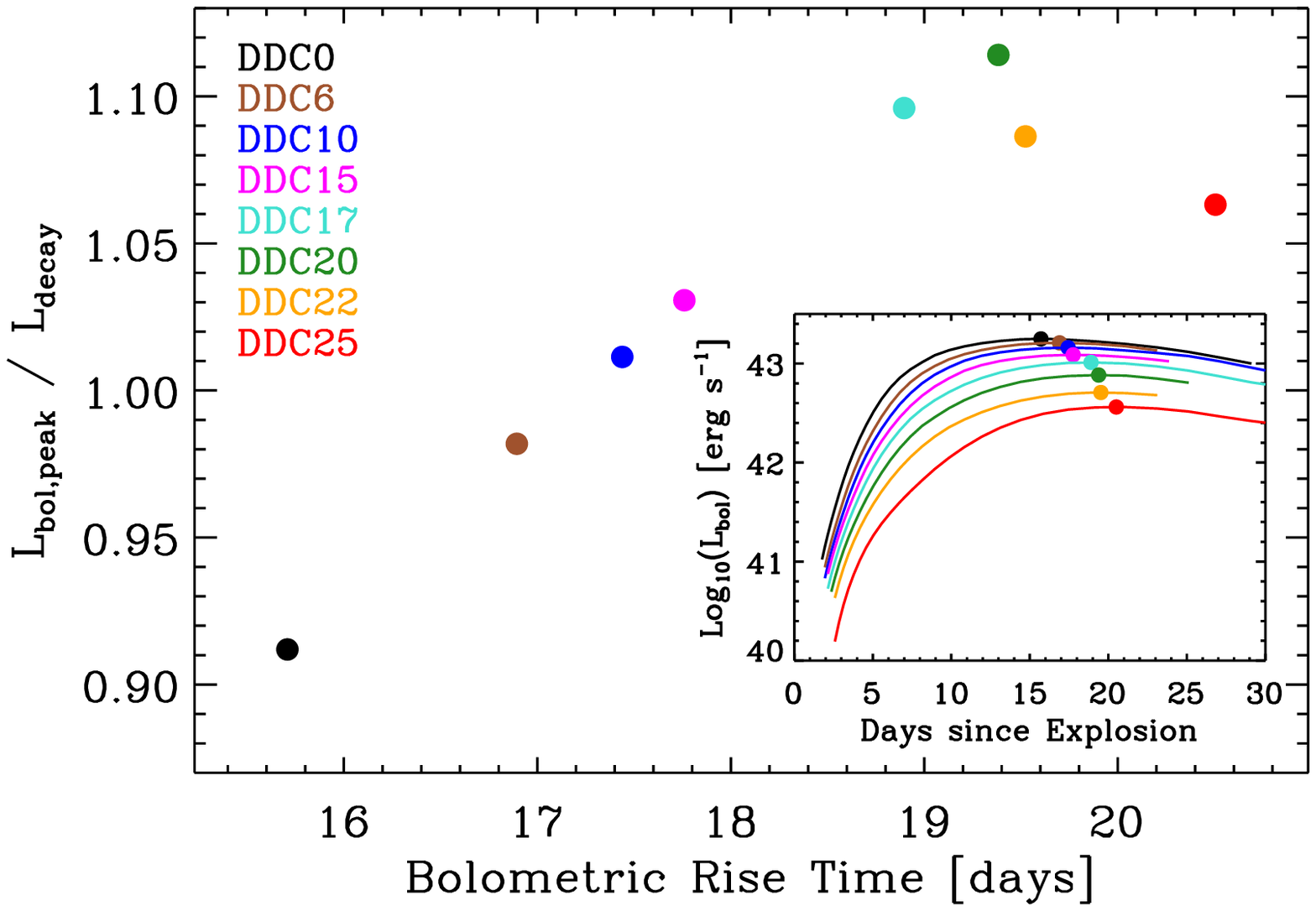,width=.475\textwidth}
\caption{\label{fig:arnett_rule}
Ratio of the peak bolometric luminosity ($L_{\rm bol,peak}$) to the
instantaneous rate of decay energy ($L_{\rm decay}$) versus rise time
to bolometric maximum for our model set (see also
\citealt{Khokhlov/etal:1993}, their Fig.~29). Our results suggest that 
Arnett's rule ($L_{\rm bol,peak}/L_{\rm decay}=1$) is
accurate to within 10\% for such delayed-detonation models.
The inset shows the bolometric light curves up until a few days
past bolometric maximum (marked with a filled circle).
}
\end{figure}

For all models the ratio of the peak bolometric luminosity to 
the instantaneous rate of decay energy (noted $Q_\gamma$ in
Table~\ref{tab:ejectaprop}) is within $\sim$\,10\% of unity
(Fig.~\ref{fig:arnett_rule}), thus supporting the so-called ``Arnett
rule'' \citep{Arnett:1979,Arnett:1982a}. This rule proposes that the
light curves of the bolometric luminosity and the instantaneous decay
rate should cross around the time of bolometric maximum. 
At this time, the ejecta Rosseland-mean 
optical depth in our simulations is a few tens, so that there is still
trapped radiation within the SN, making the optical light curve exceed
the $\gamma$-ray light curve after peak. It is only at nebular times
that the SN radiation is directly tied to the instantaneous energy
deposition rate.
A fraction of the decay energy will inevitably escape the
ejecta, even as early as maximum light. In our model series, the
energy actually {\it deposited} in the ejecta is 85\%--95\% of the
decay energy, resulting in a 4\%-17\% higher effective $Q_\gamma$
(noted $\widetilde{Q}$ in \citealt{Hoeflich/Khokhlov:1996}).

We also obtain a fair agreement with the delayed-detonation model
results of \citet{Hoeflich/Khokhlov:1996} for the peak bolometric
luminosities, but our rise times are 3--8\,d longer for a given
\nifs\ mass and show a systematic increase for lower \nifs\ mass. The
origin of the differences may be the opacity (and its associated
impact on optical depth and the diffusion time), whose influence has
been studied, e.g., by
\cite{Khokhlov/etal:1993,Pinto/Eastman:2000a,Pinto/Eastman:2000b}.
Since the pre-peak phase is not the focus of
this study, we defer a detailed discussion of this result, and
comparisons to alternate works, to Blondin et al. (in prep).

\begin{figure}
\centering
\epsfig{file=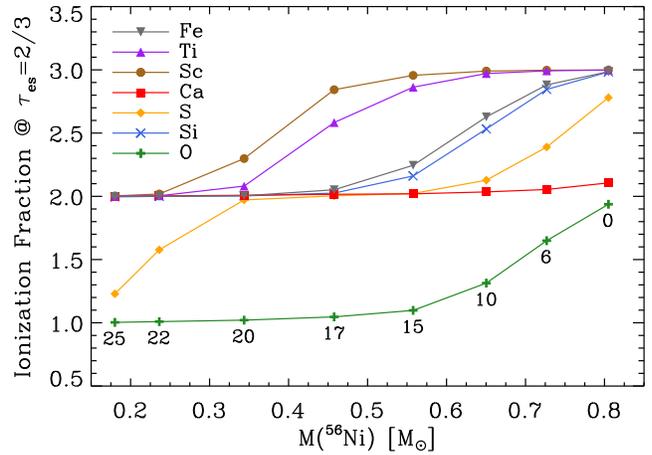,width=.475\textwidth}
\caption{\label{fig:comp_peak_phot_ionfrac}
Ionization fraction at bolometric maximum for
selected species as a function of \nifs\ mass, at the location where
the inward-integrated electron scattering optical depth $\tau_{\rm
  es}=2/3$. 
We define the ionization fraction as $\sum_i{iX^{i+}}/\sum_i{X^{i+}}$, where
$X^{i+}$ is the mass fraction of ionization stage $i$ for element
$X$. Numbers below the curve for the oxygen ionization fraction
correspond to the DDC model number.
} 
\end{figure}

At bolometric maximum, the ``photospheric'' radii\footnote{In SNe Ia,
there is no clearly-defined photosphere due the ubiquitous presence
of lines and the sub-dominance of absorption/continuum processes.
Hence, by this term, we broadly refer to the spectrum-formation
region, and adopt for that region the location where the
inward-integrated electron-scattering optical depth is 2/3.}  are on
the order of 
$\sim$\,1.3--1.4$\times$10$^{15}$\,cm irrespective of the model, so
that the range of luminosities reflects a modulation in temperature,
from 9200\,K in model DDC25 to 15270\,K in model DDC0 at this
location.  Consequently, a 
trend of higher ejecta ionization is obtained for models with
increased \nifs\ (Fig.~\ref{fig:comp_peak_phot_ionfrac}).  
Combined with abundance variations in the
spectrum formation region, these variations are at the origin of the
basic diversity of SN~Ia maximum-light spectra
(Sect.~\ref{sect:specprop}).

\begin{figure*}
\centering
\epsfig{file=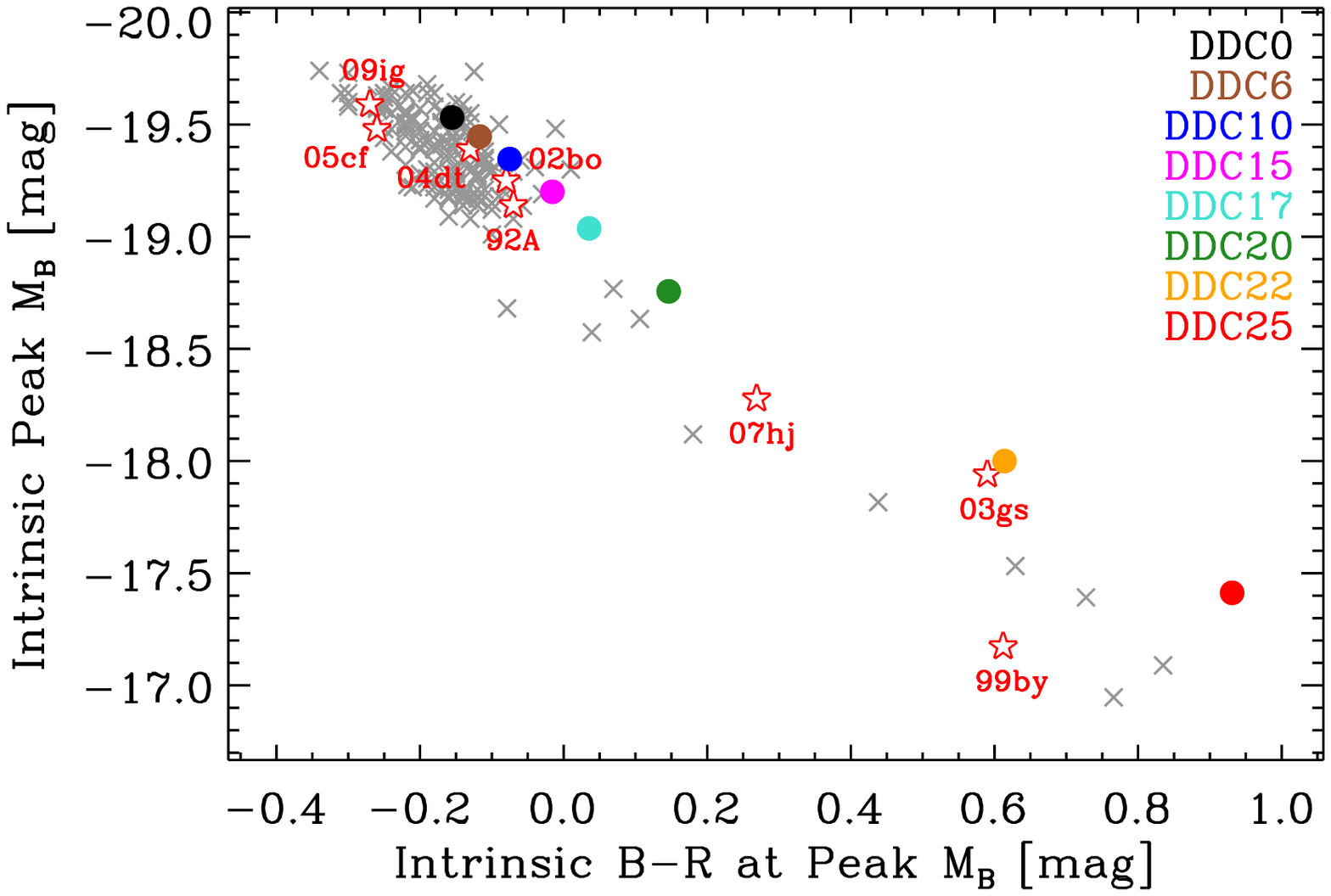,height=5.75cm}\hspace{.75cm}
\epsfig{file=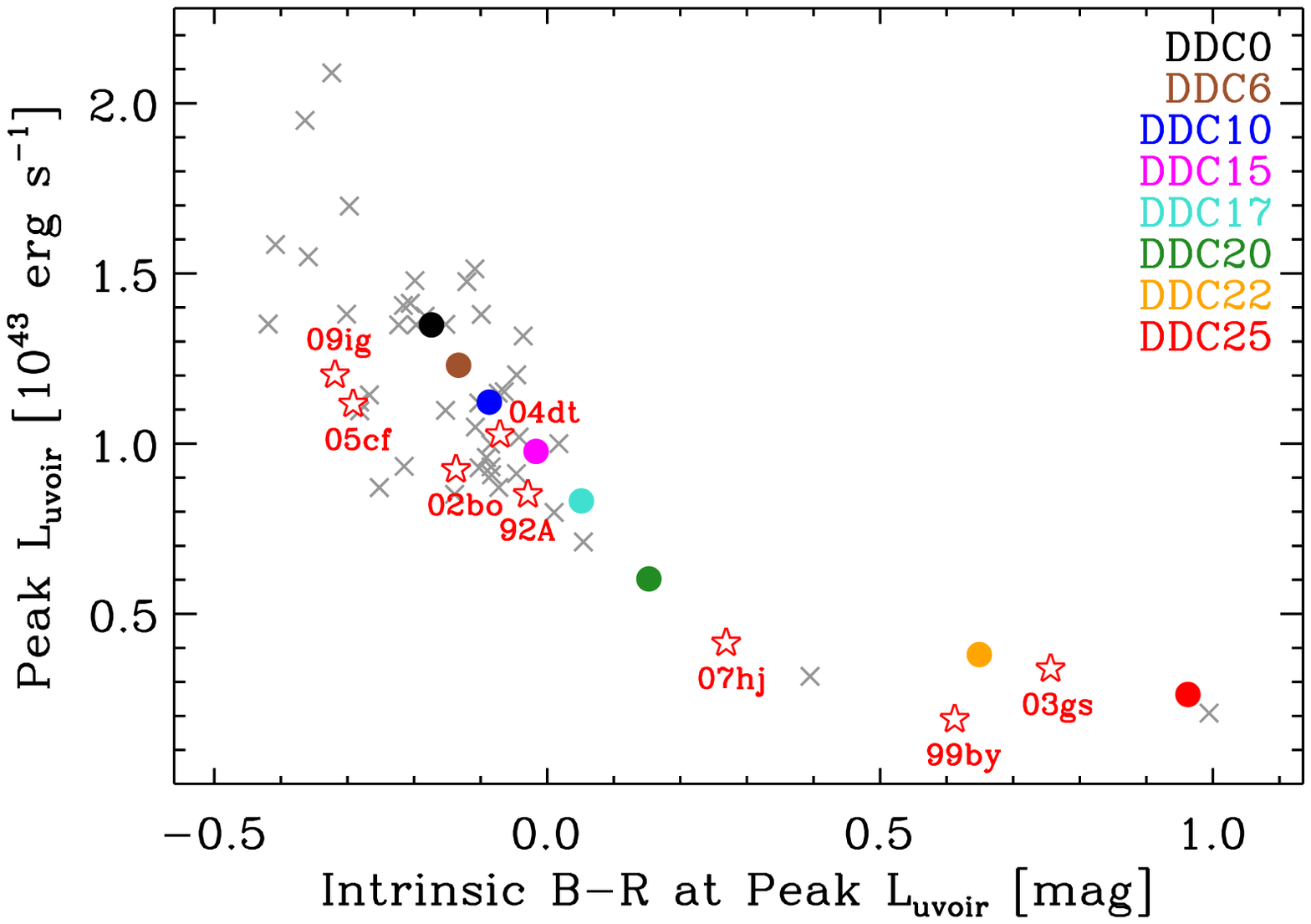,height=5.75cm}
\caption{\label{fig:comp_mag}
Intrinsic peak $M_B$ (left) and UVOIR luminosity (right) vs. intrinsic
$B-R$ colour 
at $B$-band and UVOIR maximum, respectively, for our sequence of delayed-detonation models. 
We overplot (grey crosses) the observed values 
using a wide sample of SNe Ia (similar to that used in
\citealt{Blondin/etal:2011}).
Star symbols correspond to objects that will be discussed individually
and in greater detail in Sect.~\ref{sect:spec_comp}. 
Synthetic and observed UVOIR luminosities were derived from
$UBVRI$ magnitudes (only $BVRI$ in the case of SN~2007hj) using the
method of {\protect\cite{Valenti/etal:2008}}.
We provide tabulated values of the bolometric luminosity and $UBVRIJHK$
magnitudes at bolometric maximum for our models in Table~C1.
 }
\end{figure*}

With our sample of delayed-detonation models, which spans a range in
\nifs\ mass from 0.18 to 0.81\,\msun, we reproduce the full range of
observed luminosities, $B$-band magnitudes and colours of SNe Ia at
maximum light (Fig.~\ref{fig:comp_mag}).  For example, our model set
covers $B-R$ values from $-0.17$ to 0.96\,mag at bolometric maximum
(Table~\ref{tab:ejectaprop}), and misses only a few of the brightest
and bluest \sneia\ at peak (most of these are 91T-like).  We also
reproduce the small luminosity scatter in the near-infrared bands with
respect to the optical (see Table~C1), as first
found observationally by \cite{Krisciunas/etal:2004a}.  The remarkable
achievement of the delayed-detonation mechanism, which is not new, is
that a single parameter can explain the most fundamental properties of
SN~Ia light curves.  In the next section, we show how well it
reproduces the fundamental spectroscopic properties of a wide range of
\sneia\ at bolometric maximum.

\section{Spectroscopic properties at bolometric maximum}
\label{sect:specprop}


\subsection{Spectral Diversity, ionization sequence, and line blanketing}\label{sect:spec_overall}

\begin{figure*}
\centering
\epsfig{file=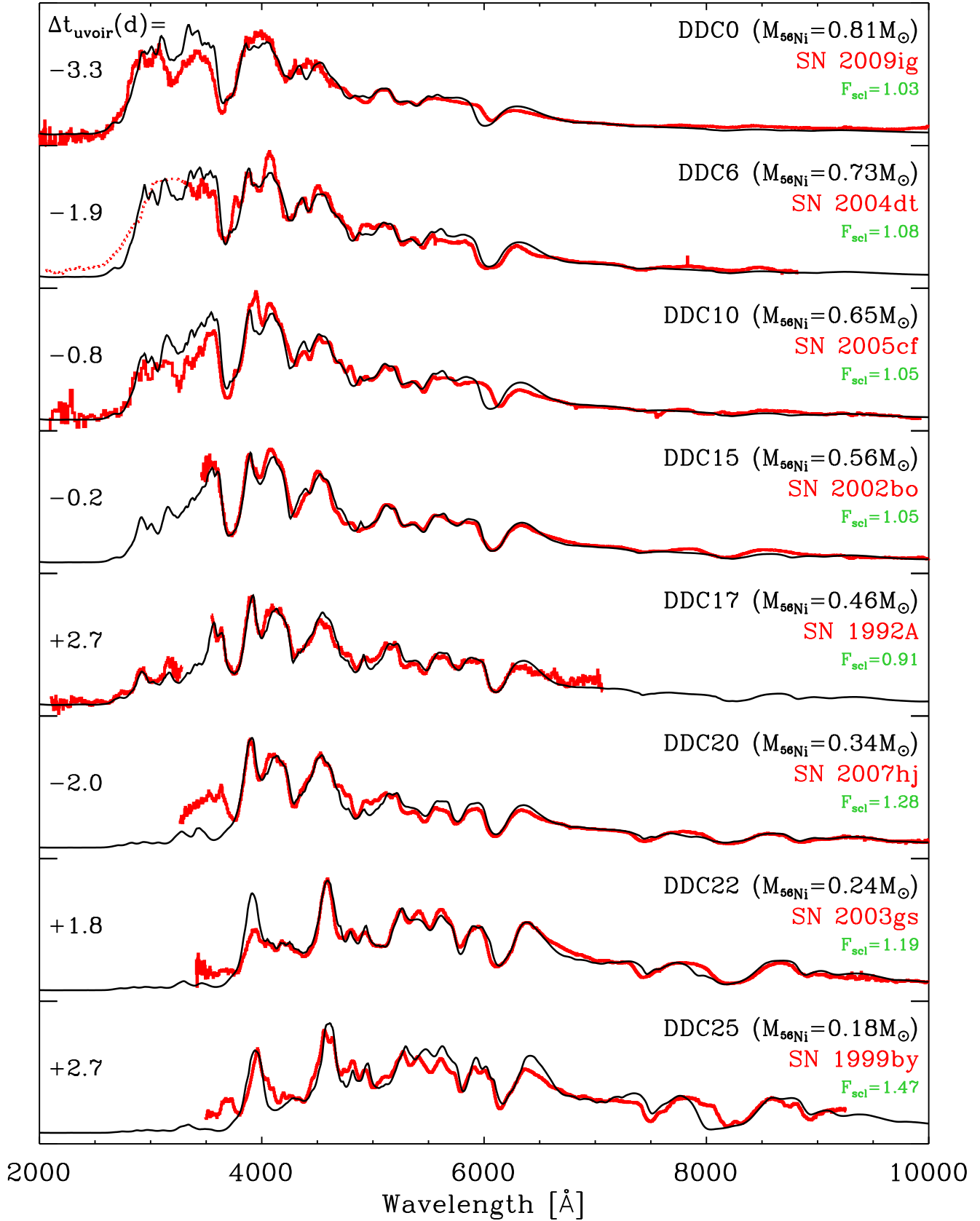,width=.95\textwidth}
\caption{\label{fig:comp_maxspec}
Comparison of synthetic spectra of delayed-detonation models (black)
to observed \sneia\ near maximum light (red).  The tickmarks on the
ordinate give the zero-flux level.  The time label corresponds to days
from pseudo-bolometric (UVOIR) maximum.  The observed spectra have
been de-redshifted, de-reddened, and scaled to match the absolute
$V$-band magnitude inferred from the corresponding
\snia\ photometry. An additional scaling ($F_{\rm scl}$; green label)
has been applied to the observed spectra to reproduce the mean
synthetic flux in the range 4000-6000\,\AA. For SN~2004dt we add the
low-resolution $HST$ ACS grism spectrum (dotted line) published by
{\protect\cite{WangX/etal:2012}}.
}
\end{figure*}

The spectral diversity of these delayed-detonation models, associated
with variations in ejecta characteristics (temperature, ionization) and
photometric properties (luminosity, colour) reflects closely the
observed diversity of \sneia\ at maximum light
(Fig.~\ref{fig:comp_maxspec}; see also Sect.~\ref{sect:spec_comp}).
This is a real achievement given the much greater complexity of
spectra compared to photometry (for example, a given model may
reproduce the colour of a \snia\ but may match the spectrum poorly;
see, e.g., \citealt{Blondin/etal:2011}).  The theoretical
spectral-energy distributions (SEDs) match satisfactorily the entire
observed wavelength range from the blue, luminous, but
spectroscopically normal SN~2009ig (model DDC0), to the redder,
low-luminosity 91bg-like SN~1999by (model DDC25).  For each
intermediate model, we also find suitable matches.  Models with a
larger \nifs\ mass achieve a higher peak bolometric luminosity and
bluer colours, which in spectra is best reflected with the stronger
flux shortward of $\sim$\,4000\,\AA.  All these properties directly
reflect the varying levels of temperature/ionization discussed in the
preceding section.  

With a total \nifs\ mass of 0.81\,\msun, model DDC0 matches the
peak luminosity of several 91T-like \sneia\ but fails to reproduce
their blue spectra characterized by lines of Fe\three\ and very weak 
Si\two\,6355\,\AA\ absorption.
In the future, we will investigate if a delayed-detonation model with a
(slightly) larger \nifs\ mass than DDC0 would yield a better agreement
with 91T-like objects.
  
Irrespective of \nifs\ mass, our models are systematically faint in
the UV. The total integrated flux blueward of 2500\,\AA\ represents
between $\sim$\,0.6\% (model DDC0) and $<0.01$\% (models DDC22 and
DDC25) of the bolometric flux. This agrees with the few observations
shown in Fig.~\ref{fig:comp_maxspec}, a subset of which have UV
photometry obtained with the {\it Hubble Space Telescope} (SN~1992A,
\citealt{Kirshner/etal:1993}; SN~2004dt, \citealt{WangX/etal:2012})
and the {\it Swift} satellite (SN~2005cf, \citealt{WangX/etal:2009a};
SN~2009ig, \citealt{SN2009ig}).  In addition, the inferred $uvw2-v$
colours\footnote{The $uvw2$ and $v$ filters have a central wavelength
  of 1941\,\AA\ and 5441\,\AA, respectively \citep{Brown/etal:2010}.}
range between $\sim2.5$\,mag (DDC0) and $\sim5$\,mag (DDC25), in line
with those inferred by \citep{Milne/etal:2010} using {\it Swift}
observations of a large sample of \sneia.

The key spectroscopic signatures of \sneia\ are reproduced, such as
the near-constant strength of the distinctive Si\two\,6355\,\AA\ line
in observations of low to high luminosity \sneia.
In contrast, the relative depth at maximum absorption of
the Si\two\,5972\,\AA\ line is a strongly varying function of peak
luminosity and colour (see Sect.~\ref{sect:rsi}).
We also predict the presence at maximum light of the
Ca\two\ near-infrared triplet (8498, 8542, and 8662\,\AA) and the
emergence of a strong O\one\,7773\,\AA\ line for SNe Ia with less than
$\sim$\,0.4\,\msun\ of \nifs.  Both features are discussed in greater
detail in Sections~\ref{sect:oi} and \ref{sect:ca2}.

While the colour of the SED is in part set by the
temperature/ionization of the spectrum formation region, another
critical ingredient is line blanketing. At the peak of the light
curve, the bulk of the radiation emerges from regions where
the ejecta composition is dominated by IMEs and IGEs (see
Table~\ref{tab:ejectaprop} and Fig.~\ref{fig:comp_elem_distrib}),
whose associated line opacity supersedes the contributions from
electron scattering and continuum processes.
These different opacity sources may overlap but often do not, so that
they effectively blanket the entire spectrum, leaving no gaps for
radiation to escape except at much longer wavelengths. Hence,
blanketing induces significant redistribution of the flux to longer
wavelengths through absorption and fluorescence.

\begin{figure*}
\centering
\epsfig{file=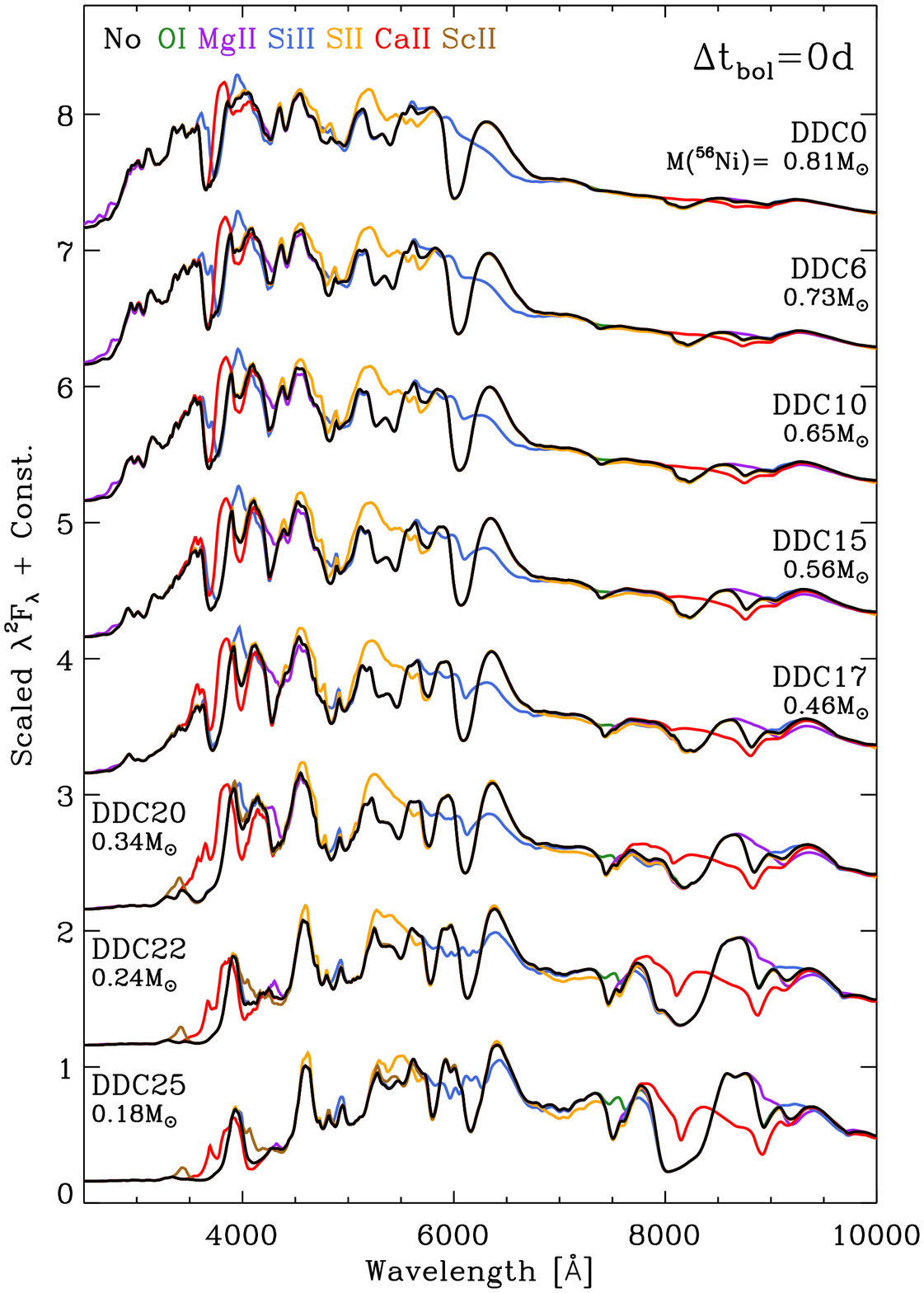,width=.475\textwidth}\hspace*{.75cm} 
\epsfig{file=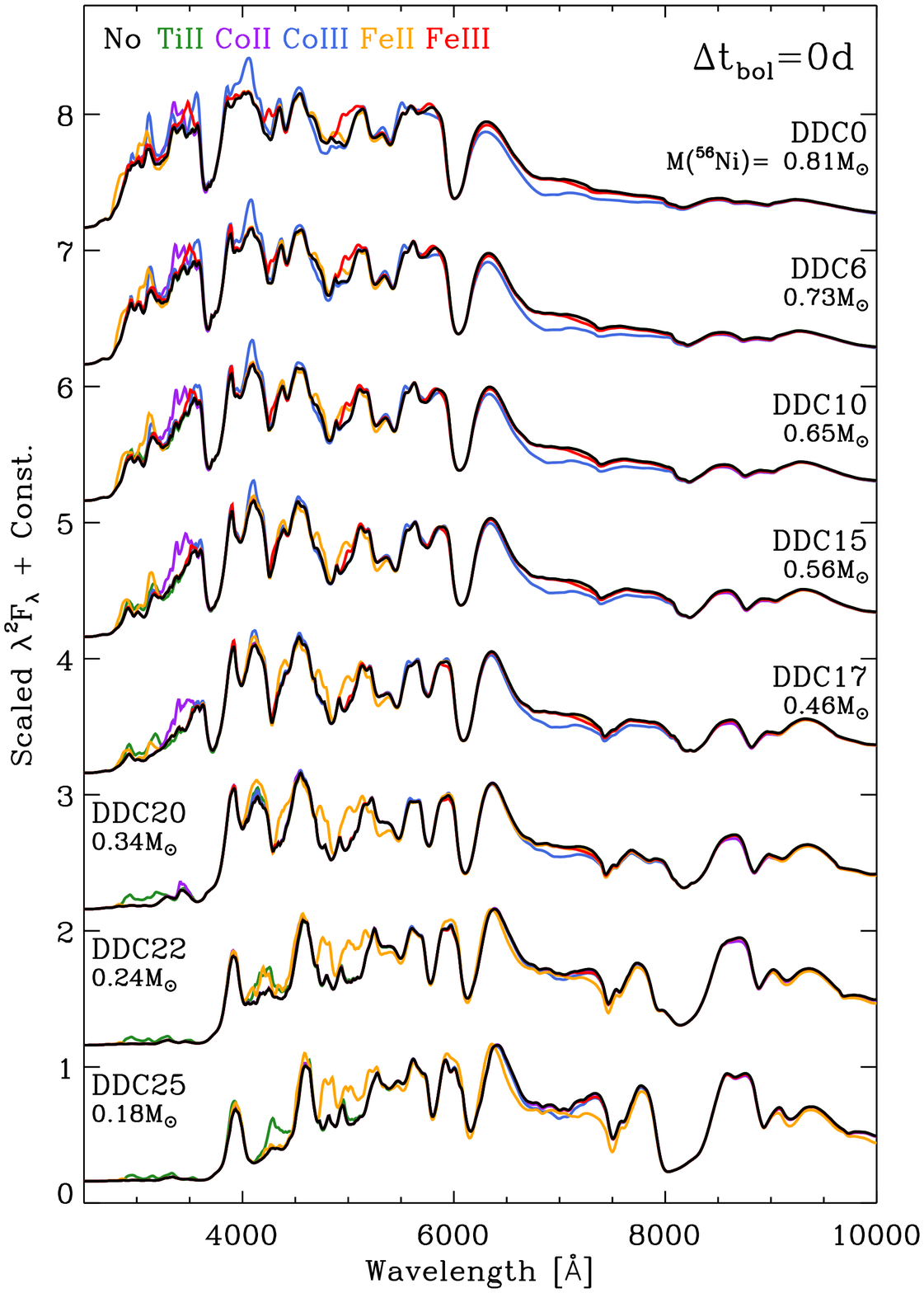,width=.475\textwidth}
\caption{\label{fig:comp_model_spec}
Comparison of synthetic spectra in our model series at
bolometric maximum illustrating individual ion contributions to
spectral features.  Besides the total spectrum (black), the left panel
shows the influence of O\one, Mg\two, Si\two, S\two, Ca\two, and
Sc\two; the right panel shows that of Ti\two, Co\two, Co\three,
Fe\two, and Fe\three\ --- the coloured lines correspond to synthetic
spectra computed {\it ignoring} all bound-bound transitions of the
corresponding ion. For better visibility of weak flux variations in
the red, we show the quantity $\lambda^2 F_{\lambda}$.
}
\end{figure*} 

Figure~\ref{fig:comp_model_spec} illustrates the impact of
lines on the maximum-light spectra and suggests two types of
influence.  First, Si\two, Ca\two, Mg\two, or O\one\ produce few
strong and isolated lines through transitions between low lying
states, which affect a limited wavelength range, although for
Ca\two\,H\&K, the width of the absorption can be as broad as the
$U$-band filter (model DDC20; see
Fig.~B6)!  Second and more dramatic
is the presence of forests of lines from some species/ions, which
effectively blanket the SN radiation.  For example, S\two\ causes
significant blanketing in the 5500\,\AA\ region (the effect is largest
for models DDC15--DDC17) --- the associated
lines overlap and prevent the identification of any of them
individually (although the strongest transitions produce the
characteristic ``W'' feature; see
Appendix~\ref{sect:ladder}). Ionization also tunes the effects of Fe
opacity, which is primarily due to Fe\three\ at high \nifs\ mass and
to Fe\two\ for models less luminous at peak than DDC15.

More striking still is the sudden increase of Ti\two\ and
Sc\two\ opacity in models with $\lesssim0.4$\,\msun\ of
\nifs\ (DDC20--DDC25), due to a shift to a lower ionization state
(Fig.~\ref{fig:comp_peak_phot_ionfrac}). This largely explains the
progressive drop in the emergent flux blueward of $\sim4500$\,\AA\ in
our synthetic spectra, which mirror the observations, as we progress
in the model sequence from DDC0 to DDC25, i.e., from blue luminous SNe
Ia to red low-luminosity 91bg-like \sneia. This effect may seem
surprising given the low abundance of Ti and Sc in the ejecta
(Table~\ref{tab:modinfo}).  The distribution of Ti is quite uniform 
throughout the ejecta with a mass fraction of $\sim10^{-5}$ at
$\tau_{\rm es}=2/3$ and beyond, such that the increasing impact of 
Ti\two\ is entirely attributed to an ionization effect. In addition,
abundance variations play a role for the Sc\two\ lines, since the Sc
mass fraction is $\sim$\,10$^{-10}$ at $\tau_{\rm es}=2/3$ (see
Table~\ref{tab:ejectaprop}), but reaches $\sim10^{-5}$ at larger
velocities. The influence of Ti\two/Sc\two\ opacity becomes
particularly severe starting at model DDC22, which marks a transition
to 91bg-like spectra, characterized by enhanced flux blocking over the
range 4000--4500\,\AA.  Note how this transition occurs over a very
narrow range in \nifs\ mass (from 0.34\,\msun\ for DDC20 to
0.24\,\msun\ for DDC22), such that in principle a strict upper limit
on the \nifs\ mass can be inferred from spectra with a 91bg-like
morphology (in the context of a Chandrasekhar-mass delayed-detonation
model).


\subsection{The Si\two\,6355\,\AA\ line and the \rsi\ ratio}
\label{sect:rsi}

The defining spectroscopic feature of \sneia, namely the blueshifted 
Si\two\,6355\,\AA\ doublet absorption, remains broad (FWHM$\sim$\,12000\,\kms) 
in models DDC0--DDC17 and only becomes noticeably narrower from DDC20 onwards, 
with FWHM$\sim$\,8000\,\kms\ in model DDC25, reflecting the
stronger confinement of Si to the inner ejecta in models with a lower
deflagration-to-detonation transition density
(Fig.~\ref{fig:comp_elem_distrib}). The velocity at maximum absorption
($v_{\rm abs}$) approximately reflects that of the peak Si abundance 
($X_{\rm Si}\sim$\,0.6, regardless of the \nifs\ mass;
see diamonds in Fig.~\ref{fig:comp_elem_distrib}), and varies smoothly
between $\sim$\,$-$17000\,\kms\ for DDC0 and $\sim$\,$-$9000\,\kms\ for
DDC25.

\begin{figure}
\centering
\epsfig{file=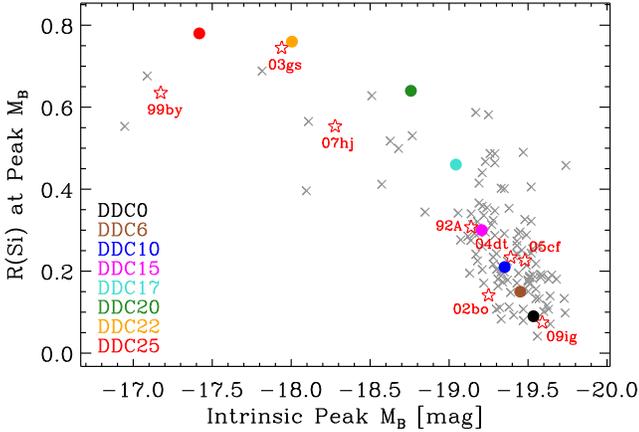,width=.475\textwidth}
\caption{\label{fig:rsi}
Spectroscopic ratio \rsi\ versus intrinsic
peak $M_B$ for our model set. We overplot (grey crosses) observed values from a 
large sample of \sneia\ (see
\citealt{Blondin/etal:2012}). Star symbols correspond to objects that
will be discussed individually and in greater detail in
Sect.~\ref{sect:spec_comp}.
}
\end{figure}

The relative depth at maximum absorption varies little amongst
our model series, as the line is optically thick at this location in
all models  (the Sobolev optical depth for the transition at
  6347\,\AA\ varies between $\sim2$ for DDC0 and $\sim387$ for
  DDC25). This contrasts with  
the neighbouring Si\two\,5972\,\AA\ line which becomes progressively
stronger with decreasing \nifs\ mass, transitioning from optically
thin to thick ($\tau_{\rm Sob}($5979\,\AA$)>1$ at maximum absorption
for models DDC20--25) and reflecting the drop in the Si ionization
fraction (Fig.~\ref{fig:comp_peak_phot_ionfrac}; see also
  \citealt{Hachinger/etal:2008}). The relative 
behaviour of both lines is a well-known spectroscopic characteristic
of \sneia, and quantified in terms of the \rsi\ ratio
\citep{Nugent/etal:1995}. 
In our models \rsi\ varies between $\lesssim0.1$
(DDC0) and $\sim0.8$ (DDC25), in good agreement with the observed
variation (Fig.~\ref{fig:rsi}). Contrary to claims by
\cite{Garnavich/etal:2004} we find no evidence for Ti\two\ absorption
in this wavelength region that could bias the \rsi\ ratio for
low-luminosity \sneia\ (see Appendix~\ref{sect:ladder}).

\cite{Branch/etal:2006} use the pseudo-equivalent widths (pEW) of the
Si\two\,6355\,\AA\ and 5972\,\AA\ lines to 
classify \sneia\ into four subgroups, organized around the ``Core
Normal'' subgroup which occupies a central position in this
two-dimensional parameter space. \sneia\ with smaller and larger
pEW(6355\,\AA) are termed ``Shallow Silicon'' (includes 91T-like
\sneia) and ``Broad Line'', respectively, while those with larger
pEW(5972\,\AA) are dubbed ``Cool'' (includes 91bg-like \sneia), in
reference to their lower ejecta temperatures. Our models  belong to only
two subclasses of this  scheme, namely Broad Line
(models DDC0--17) and Cool (DDC20-25) \sneia.

The classification scheme of \cite{WangX/etal:2009b} is based on the
velocity at maximum absorption of the Si\two\,6355\,\AA\ feature, and
divides \sneia\ into Normal and High-velocity subclasses. \sneia\
with 91T-like or 91bg-like spectra (including \sneia\ with
\rsi\,$>0.5$) are not classified according to this scheme. As
noted in Sect.~\ref{sect:spec_overall} there are no 91T-like
spectra in our model series, while models DDC20--25 either have
\rsi\,$>0.5$ (DDC20) or display 91bg-like spectra (DDC22--25). The
remaining models (DDC0--17) would all be part of the High-velocity
subclass (i.e., $v_{\rm abs}\lesssim-12000$\,\kms; see, e.g.,
\citealt{Blondin/etal:2012}).

There are thus no Core Normal \citep{Branch/etal:2006} or Normal
\citep{WangX/etal:2009b} \sneia\ amongst our model set. The
Si\two\,6355\,\AA\ line in our synthetic spectra is systematically
broader and more blueshifted than observed in these two subclasses,
suggesting the ejecta kinetic energy in our models is too large to
account for the bulk of \sneia\footnote{The simplified reduced
  nuclear kinetics scheme used in our models (see
  Sect.~\ref{sect:ddt}) may lead to an overestimation of the
  asymptotic kinetic energy. Hydrodynamical simulations with detailed
  nuclear kinetics are planned to quantify the difference.}.
However, the impact is often only visible
in the Si\two\ lines (4130\,\AA, 5972\,\AA, and 6355\,\AA), while
other spectroscopic features are well matched (e.g., Ca\two\,H\&K and
the S\two\ ``W'' feature in the comparison between model DDC10 and
SN~2005cf in Fig.~\ref{fig:comp_maxspec}), such that some of our
models closely resemble Core Normal \sneia\ despite the
Si\two\,6355\,\AA\ mismatch (DDC0 and DDC10). If this were solely a
problem related to the ejecta kinetic energy then {\it all} lines
should be too broad. Our limitation to 1D also overlooks potential
viewing-angle effects in an aspherical ejecta, which can significantly
impact the width and blueshift of the Si\two\,6355\,\AA\ line, as
shown by \cite{Blondin/etal:2011} based on the 2D delayed-detonation
models of \cite{KRW09}.


\subsection{The O\one\,7773\,\AA\ line as an abundance tracer}
\label{sect:oi}

The evolution in strength and morphology of the O\one\,7773\,\AA\
line profile in our synthetic spectra reproduces well the observed
sequence. This line becomes progressively stronger
with decreasing \nifs\ mass along our model sequence
(Fig.~\ref{fig:comp_model_spec}, left; green line).
It is essentially
non-existent in models DDC0-10, barely noticeable in models DDC15 and
DDC17, only to strongly imprint the spectra from model DDC20 onwards. This
evolution follows the gradual decrease in the
ionization fraction for O from $\sim2.0$ to 1.0
(Fig.~\ref{fig:comp_peak_phot_ionfrac}), but seems primarily influenced by
the varying oxygen abundance with \nifs\ mass (factor $\sim5$
variation in total oxygen mass between DDC0 and DDC25)
and the resulting chemical
stratification in velocity. The velocity at which $X({\rm
  O})=10^{-3}$ drops from $\sim17500$\,\kms\ in model DDC0 to
$\sim9200$\,\kms\ in model DDC25, while the corresponding
``photospheric'' velocities are 9699 and 7338\,\kms, respectively
(Table~\ref{tab:ejectaprop}). 
We thus associate this behavior of the O\one\ line with an abundance effect 
resulting from the ejecta stratification in velocity.

The O\one\,7773\,\AA\ line can in principle be used as a probe of
left-over unburnt oxygen. However, unlike C\two\ lines which provide
an unambiguous signature of unburnt carbon, the only definite
signature of unburnt oxygen (as opposed to O produced via carbon burning)
in our models is through the detection of absorption out to velocities
where we recover the initial $X({\rm O})\approx0.5$ composition. The minimum
velocity at which $X({\rm O})\approx0.5$ varies between
$\sim20000$\,\kms\ (model DDC25) and $\sim35000$\,\kms\ (model DDC0),
and we obtain no O\one\,7773\,\AA\ absorption out to these velocities
in any of our models.
This is likely a shortcoming of spherical
symmetry, as multi-D effects are considered essential to preserve O (and
C) unburnt in the ejecta, even at the relatively low velocities
corresponding to the spectral-formation region at bolometric maximum
\citep[see][]{Gamezo/Khokhlov/Oran:2005}. 


\subsection{Impact of Ca\two\ lines and high-velocity features}
\label{sect:ca2}

\begin{figure}
\centering
\epsfig{file=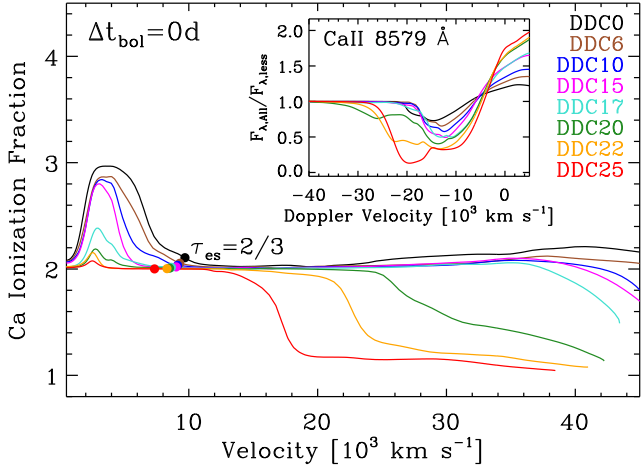,width=.475\textwidth}
\caption{\label{fig:comp_model_ionfrac}
Ionization fraction for Ca at bolometric maximum as a function of
velocity. The solid dots indicate the location where the
inward-integrated electron scattering optical depth $\tau_{\rm
  es}=2/3$. The inset shows Ca\two\,near-IR triplet line profiles
computed as the ratio of the full spectrum ($F_{\lambda,{\rm All}}$)
to that excluding all Ca\two\ bound-bound transitions
($F_{\lambda,{\rm less}}$), revealing high-velocity absorption
features in models with less \nifs.
} 
\end{figure}

The large drop in flux around $\sim3500$\,\AA\ between model DDC17 and
DDC20 is mainly due to an opacity increase associated with
Ca\two. While the Ca ionization fraction at $\tau_{\rm es}=2/3$ stays
fairly constant from models DDC0 to DDC25, it changes significantly at
velocities $\gtrsim
15000$\,\kms\ (Fig.~\ref{fig:comp_model_ionfrac}). Since the
Ca\two\,H\&K lines remain optically thick out to large velocities,
absorbing well beyond 30000\,\kms\ in all models, such ionization
changes have a noticeable impact on the emergent flux in the $U$ band.

The effect is more clearly visible in the Ca\two\,near-infrared
triplet, which remains weak in all models up to DDC17 and becomes
progressively stronger up to DDC25, with a line-profile morphology
displaying high-velocity components (see inset in
Fig.~\ref{fig:comp_model_ionfrac}). Such high-velocity features (HVFs)
were associated with density or abundance enhancements by
\cite{Mazzali/etal:2005b}, possibly resulting from multi-dimensional
effects in the explosion itself or from interaction with circumstellar
material.  Our simulations suggest that HVFs in Ca\two\ can even arise
in spherically symmetric ejecta, primarily through optical-depth
effects due to ionization changes occurring at large velocity.
Abundance effects play no role here: the Ca mass fraction at
15000\,\kms\ is three orders of magnitude lower in model DDC25
($\sim4.1\times10^{-5}$), which displays prominent HVFs, compared to
model DDC0 ($\sim 3.4\times10^{-2}$), which displays a single weak
absorption. Furthermore, there is no density bump at these velocities.
Obviously, when such HVFs are associated with a variation in the
polarization signal across the line (e.g., in SN~2001el;
\citealt{WangL/etal:2003}), a departure from spherical symmetry must
occur, but these two signatures (HVF and polarization) can arise
independently of one another.

In all of our models with more than 0.4\,\msun\ of \nifs\ (DDC0--17),
the Ca\two\,H\&K absorption feature is strongly contaminated by the
Si\two\,3858\,\AA\ doublet, resulting in
double-absorption features (see Appendix~\ref{sect:ladder}). These are
occasionally mistaken for ``photospheric'' and ``high-velocity''
components of Ca\two\,H\&K (see discussion in
\citealt{Lentz/etal:2000}).

\section{Individual comparisons of delayed-detonation models to
  observations at maximum light} 
\label{sect:spec_comp}

\begin{table*}
\caption{Comparison of delayed-detonation models to observed \sneia\ near maximum light.}\label{tab:comp_maxspec}
\begin{tabular}{lcccclccccccc}
\hline\hline
\multicolumn{1}{c}{Model} & $M(\nifs)$ & $t_{\rm rise}$ & $L_{\rm bol}$   & $L_{\rm uvoir}$   & \multicolumn{1}{c}{SN}  & Ref. & $\mu$ & Method & $E(B-V)$ & $Q_{\rm uvoir}$ & \multicolumn{1}{c}{$\Delta t_{\rm uvoir}$} & $F_{\rm scl}$ \\
                          &[M$_{\sun}$]& [day]          & [erg\,s$^{-1}$] & [erg\,s$^{-1}$]   &                         &      & [mag] &        & [mag]    &                 & \multicolumn{1}{c}{[day]}                &               \\
\multicolumn{1}{c}{(1)}   & (2)        & (3)            & (4)             & (5)               & \multicolumn{1}{c}{(6)} & (7)  & (8)   & (9)    & (10)     & (11)            & \multicolumn{1}{c}{(12)}                 & (13)          \\
\hline
DDC0     & 0.805 & 15.71 & 1.77(43) & 1.20(43) & 2009ig   & 1        &    32.60 &              TF &     0.14 &     1.12 &   $-$3.3 &     1.03 \\
DDC6     & 0.727 & 16.89 & 1.62(43) & 1.03(43) & 2004dt   & 2,3      &    34.44 &   $z_{\rm CMB}$ &     0.04 &     1.20 &   $-$1.9 &     1.08 \\
DDC10    & 0.650 & 17.44 & 1.45(43) & 1.12(43) & 2005cf   & 4,5      &    32.17 &            SNIa &     0.22 &     1.00 &   $-$0.8 &     1.05 \\
DDC15    & 0.558 & 17.76 & 1.22(43) & 9.25(42) & 2002bo   & 6        &    31.90 &            SNIa &     0.41 &     1.06 &   $-$0.2 &     1.05 \\
DDC17    & 0.458 & 18.90 & 1.03(43) & 8.51(42) & 1992A    & 7        &    31.63 &             SBF &     0.02 &     0.98 &     +2.7 &     0.91 \\
DDC20    & 0.344 & 19.38 & 7.65(42) & 4.15(42) & 2007hj   & 8        &    33.88 & $z_{\rm Virgo}$ &     0.11 &     1.32 &   $-$2.0 &     1.28 \\
DDC22    & 0.236 & 19.52 & 5.10(42) & 3.39(42) & 2003gs   & 9        &    31.65 &             SBF &     0.07 &     1.12 &     +1.8 &     1.19 \\
DDC25    & 0.180 & 20.50 & 3.64(42) & 1.91(42) & 1999by   & 10       &    30.74 &        Cepheids &     0.03 &     1.38 &     +2.7 &     1.47 \\
\hline
\end{tabular}

\medskip
\flushleft
{\bf Notes:} Numbers in parentheses correspond to powers of ten.
{\bf Column information:} 
(1) Model name;
(2) mass of synthesized \nifs;
(3) rise time to peak bolometric luminosity. Note that the rise to UVOIR maximum is $\sim$0.5--1\,d longer than to the true bolometric maximum;
(4) peak bolometric luminosity;
(5) peak UVOIR luminosity derived from synthetic $UBVRI$ magnitudes using the method of \citealt{Valenti/etal:2008};
(6) name of observed \snia\ used for comparison in Fig.~\ref{fig:comp_maxspec};
(7) reference for spectroscopic data used in Fig.~\ref{fig:comp_maxspec} (see References below);
(8) distance modulus (assuming $H_0=73$\,km\,s$^{-1}$\,Mpc$^{-1}$ where applicable);
(9) method for distance modulus determination: SBF=Surface Brightness Fluctuations, SNIa=SN-based distance, TF=Tully Fisher, $z_{\rm CMB}$=CMB-frame redshift-based distance modulus; $z_{\rm Virgo}$=distance modulus based on the recession velocity corrected for the influence of the Virgo cluster;
(10) total (Galactic and host-galaxy) reddening. Galactic reddening values are based on the infrared dust maps of \cite{SFD98}, while host-galaxy reddening was inferred based on our own fits to optical (and NIR when available) SN lightcurves. For SN~2003gs we use the reddening reported by \cite{Krisciunas/etal:2009};
(11) ratio of synthetic to observed peak UVOIR luminosities, derived from $UBVRI$ magnitudes (only $BVRI$ in the case of SN~2007hj) using the method of \citealt{Valenti/etal:2008};
(12) phase of SN with respect to the time of UVOIR maximum;
(13) scale factor applied to the observed spectra to match the mean synthetic flux in the wavelength range 4000--6000\,\AA. \\
{\bf References:} 
(1): \citealt{SN2009ig},
(2): \citealt{Altavilla/etal:2007},
(3): \citealt{WangX/etal:2012},
(4): \citealt{Garavini/etal:2007b},
(5): \citealt{Bufano/etal:2009},
(6): \citealt{Benetti/etal:2004},
(7): \citealt{Kirshner/etal:1993},
(8): \citealt{Silverman/etal:2012a},
(9): Kotak et al. (in prep.),
(10): \citealt{Garnavich/etal:2004} \\
\end{table*}

For each Chandrasekhar-mass delayed-detonation model in our set we are
able to find a good match to an observed
\snia\ (Fig.~\ref{fig:comp_maxspec}). In most cases there are several
observed \sneia\ that match a given model --- this is not surprising
given the overall spectral homogeneity of \sneia\ with a given
luminosity \citep[see, e.g.,][]{Blondin/etal:2012}. In those cases we
preferentially select \sneia\ whose spectra cover the largest
wavelength range (i.e. extending blueward of Ca\two\,H\&K and redward
of the Ca\two\ NIR triplet). The comparison between synthetic and
observed spectra is largely based on a ``chi-by-eye'' estimate of the
overall SED match, with particular emphasis on the flux level in the
$U$ and $B$ bands, as well as individual spectroscopic features. This
approach is fine to assess the broad compatibility of
delayed-detonation models --- given the recognized influence of
multi-D effects and our 1D approach, a perfect match may be
coincidental.  We use the Supernova Identification (SNID) code of
\cite{SNID} to ease our exploration of the largest possible data set,
but make no attempt at quantifying the strength of the correlation
between synthetic and observed spectra as done by
\cite{Blondin/etal:2011}.


\subsection{DDC0 ($M_{\rm \nifs}=0.81$\,\msun) vs. SN~2009ig}

The spectra of SN~2009ig \citep{SN2009ig} are well reproduced by DDC0
over the wavelength range 2000--10000\,\AA. 
We use the same distance modulus as \cite{SN2009ig}
(from \citealt{Tully/Fisher:1988}), but a larger extinction
($A_V=0.3$\,mag from our own fits to the light curves
using the BayeSN statistical model of \citealt{Mandel/etal:2011}).
\cite{SN2009ig} do note however the presence of significant
interstellar Na\one\ absorption (${\rm EW}=0.4$\,\AA) at the redshift
of the host galaxy, consistent with visual extinctions in the range
$0\lesssim A_V\lesssim 1$\,mag \citep[see][their Fig.~5]{Blondin/etal:2009}. 

SN~2009ig is a ``Core Normal'' \snia\ 
according to the \cite{Branch/etal:2006} classification scheme
\citep{Blondin/etal:2012}. We are thus unable to reproduce its
narrow Si\two\,6355\,\AA\ line profile (see Sect.~\ref{sect:rsi}).
There is only a minor mismatch, however, in the width of other
features, such as the Ca\two\,H\&K absorption (contaminated by
Si\two\,3858\,\AA\ in the blue; see Sect.~\ref{sect:ca2} and
Fig.~B1).
The comparison at $-3.3$\,d is impressive overall, although it is not
as good for spectra closer to maximum light, in particular in the
region 3000--3500\,\AA\ where the model is brighter than observations.
The ratio of peak UVOIR luminosities is comparable 
($Q_{\rm uvoir}=1.12$)
to the scaling to match the mean fluxes in the
4000--6000\,\AA\ region ($F_{\rm scl}=1.03$), confirming the 
good correspondence between the model and the data.

The peak luminosity and colours of DDC0 match well the luminous
SN~1999ac \citep{Garavini/etal:2005,Phillips/etal:2006} and SN~2001eh
\citep{Sauer/etal:2008}, although our synthetic line profiles
are systematically too broad (in particular Ca\two\,H\&K, in addition
to Si\two\,6355\,\AA) with respect to the observations. 

Model DDC0 also reproduces the overall SED and spectroscopic
features of the luminous yet spectroscopically normal SN~1999ee
\citep{Hamuy/etal:2002}, and matches well its peak luminosity ($Q_{\rm
uvoir}=0.98$), although the flux level blueward of $\sim3500$\,\AA\
is too high in our synthetic spectrum.


\subsection{DDC6 ($M_{\rm \nifs}=0.73$\,\msun) vs. SN~2004dt}

SN~2004dt has some interesting properties. Its Si\two\,6355\,\AA\ line
displays prominent HVFs up until a few days past maximum light, which
led \cite{Altavilla/etal:2007} to invoke an enhanced density
structure in the outer layers.
Furthermore, it displays a UV excess around maximum light compared to
other \sneia, which led \cite{WangX/etal:2012} to invoke
viewing-angle effects associated with large-scale ejecta asymmetries.
Interestingly, our smooth 1D model DDC6 reproduces the line widths, 
the overall SED, and the strong UV
flux of SN~2004dt. (The dotted line in Fig.~\ref{fig:comp_maxspec}
shows an $HST$ UV grism spectra taken with the ACS instrument --- the
resolution is very low so we are not concerned with small-scale
absorption features.) 

We find an excellent match to the peak luminosity, colours, and
early-time spectra of SN~2006cp \citep{Blondin/etal:2012}, 
although the published spectra span a limited wavelength range 
(3500--7500\,\AA) and none are within 5\,d from $B$-band maximum light.


\subsection{DDC10 ($M_{\rm \nifs}=0.65$\,\msun) vs. SN~2005cf}

SN~2005cf is a ``golden standard'' amongst \sneia\ according to
\cite{WangX/etal:2009a}.  The match of the maximum-light spectra to
model DDC10 is excellent overall in the range 2000--10000\,\AA,
although the Si\two\ lines (4130\,\AA, 5972\,\AA, and 6355\,\AA) are
significantly narrower and less blueshifted than in our synthetic
spectrum (as expected since this SN is part of the Core Normal
subclass; see Sect.~\ref{sect:rsi}). The widths of other features,
such as Ca\two\,H\&K and the S\two\ ``W'', are nonetheless compatible
with our model. The Ca\two\,H\&K absorption displays a
double-absorption profile whose blue component is mostly due to
contamination by Si\two\,3858\,\AA\ (Sect.~\ref{sect:ca2} and
Fig.~B3).  Model DDC10 also
reproduces the UV flux for this SN, measured using the {\it Swift}
satellite, and matches perfectly the observed UVOIR flux ($Q_{\rm
  uvoir}=1.00$).

The broader Si\two\ lines in the spectra of SN~1997bp
\citep{Blondin/etal:2012} are better matched by DDC10, but this SN
only has published spectra in the range 3500--7500\,\AA.


\subsection{DDC15 ($M_{\rm \nifs}=0.56$\,\msun) vs. SN~2002bo}

There are only SN-based distances to SN~2002bo, which vary by
$\sim0.5$\,mag in distance modulus (i.e. $\sim20$\% in absolute
distance). This uncertainty stems in part from the large reddening
inferred for this SN ($E(B-V)\approx0.4$\,mag, the largest of our
comparison sample; see Table~\ref{tab:comp_maxspec}). We adopt the 
distance modulus of 31.90\,mag inferred from NIR light curves
\citep{Wood-Vasey/etal:2008}, less subject to extinction than the
visual bands. A similar distance modulus is derived from NIR light
curves of the highly-reddened SN~2002cv ($\mu=31.85$\,mag assuming
$H_0=73$\,km\,s$^{-1}$\,Mpc$^{-1}$; \citealt{Wood-Vasey/etal:2008}),
which occurred in the same host galaxy as SN~2002bo. Our inferred
host-galaxy reddening ($E(B-V)=0.39$\,mag from our own fits to the
lightcurves using the BayeSN statistical model of
\citealt{Mandel/etal:2011}) is consistent with published estimates
\citep[e.g.,][]{Benetti/etal:2004}. The authors of this latter paper
argue based on spectroscopic modelling that a lower value of the
reddening ($E(B-V)\approx0.30$\,mag) is needed to match the
maximum-light spectrum, although their synthetic spectra
systematically overestimate the flux redward of $\sim6500$\,\AA\ due
to their assumption of an inner diffusive boundary (their Fig.~17).

The synthetic spectrum of model DDC15 and the observations of
SN~2002bo nearly overlap. The peak UVOIR luminosity is well reproduced
($Q_{\rm uvoir}=1.06$, i.e. better than 10\%), as are the magnitudes
at bolometric maximum in the optical and NIR bands (see
Table~C1).  

We obtain equally-good matches to SN~2002dj \citep{Pignata/etal:2008},
SN~2004fu, and SN~2007co (both with spectra published by
\citealt{Blondin/etal:2012}), all of which are also Broad Line
\sneia\ of comparable luminosity.

A decent match is also found to the nearby SN~2011fe
\citep{Nugent/etal:2011}, although our model DDC15 is significantly
redder at maximum light (but of comparable luminosity, with $Q_{\rm
  uvoir}\approx 1.1$). \cite{Roepke/etal:2012} found a reasonable
match to SN~2011fe using a 3D delayed-detonation model tuned to
synthesize 0.6\msun\ of \nifs, although they argue that the agreement
at maximum light is better for their violent WD merger model.

DDC15 is also a good match to SN~2001ay, for which
\cite{Baron/etal:2012} invoke a pulsating delayed detonation 
model with a special WD configuration (excess C abundance and low
central density). Our standard delayed-detonation model DDC15 is able to reproduce the
peak UVOIR luminosity ($Q_{\rm uvoir}=1.05$) and colours, as well as
the broad spectroscopic features (in particular
Si\two\,6355\,\AA).  


\subsection{DDC17 ($M_{\rm \nifs}=0.46$\,\msun) vs. SN~1992A}

\cite{Kirshner/etal:1993} present an extensive set of UV spectra
for SN~1992A taken with the International Ultraviolet Explorer ({\it
  IUE}) as well as {\it HST}. The maximum-light spectrum shown in
Fig.~\ref{fig:comp_maxspec} is a combination of an {\it IUE} UV
spectrum and a CTIO optical spectrum. \cite{Kirshner/etal:1993} obtain
reasonable fits to the observed spectra using a parameterized
steady-state LTE radiative transfer calculation of the
delayed-detonation model DD4 of \cite{Woosley:1991}.
Our full non-LTE time-dependent calculation reproduces the
spectrum over the entire
observed wavelength range, and is consistent with the inferred peak
UVOIR luminosity ($Q_{\rm uvoir}=0.98$).

SN~2004eo (labeled ``transitional'' by
\citealt{Pastorello/etal:2007b}) is also a fair match to model DDC17, 
although $Q_{\rm uvoir}\approx0.7$, i.e. model DDC17 is fainter. Yet
its \nifs\ mass (0.46\msun) is $\sim0.1$\,\msun\ larger than that
inferred by \cite{Mazzali/etal:2008} based on modelling a single
nebular-epoch spectrum (0.38\,\msun). However, \cite{Mazzali/etal:2008} 
note that their inferred \nifs\ mass falls short of
reproducing the peak bolometric luminosity, and revise their estimate
to $0.43\pm0.05$\,\msun, in agreement with model DDC17. 
A clear asset of our approach is the simultaneous calculation of the
light curves and the spectra, which ensures the overall spectral
morphology accurately reflects the SN luminosity.


\subsection{DDC20 ($M_{\rm \nifs}=0.34$\,\msun) vs. SN~2007hj}

Model DDC20 corresponds to a transition from standard to
low-luminosity \sneia, with the noticeable impact of increasing
blanketing from Ti\two/Sc\two\ around $\sim4000$\,\AA.

A good match is found to SN~2007hj. This SN is slightly bluer than the
model but the relative strengths of spectral features is well
reproduced, in particular the 4000--4500\,\AA\ range, the S\two\ ``W''
feature, the Si\two\ lines (5972\,\AA\ and 6355\,\AA) and
corresponding \rsi\ ratio, the O\one\,7773\,\AA\ absorption, the
Ca\two\ near-IR triplet, as well as the Mg\two\ feature at
$\sim8800$\,\AA\ (see Fig.~B6). There
are however significant differences in the blue wing of the
Ca\two\,H\&K line which is strongly suppressed in the model mainly due
to Ca\two\ absorption out to high velocities (see
Sect.~\ref{sect:ca2}). Lines of Ti\two\ and Sc\two\ also deplete the
flux in this region.

The spectroscopic features of SN~1998bp \citep{Matheson/etal:2008} and
SN~2007au \citep{Blondin/etal:2012} are well matched by model DDC20,
but the predicted luminosity is too high ($Q_{\rm uvoir}=1.91$ and
1.62, respectively). The match to SN~2000dk \citep{Matheson/etal:2008}
is of comparable quality as the match to SN~2007hj, although the
discrepancy in the blue wing of Ca\two\,H\&K is even larger (the
luminosity is however in better agreement, with $Q_{\rm uvoir}=0.85$).


\subsection{DDC22 ($M_{\rm \nifs}=0.24$\,\msun) vs. SN~2003gs}

With DDC22 we enter the category of subluminous \sneia, with the
characteristic broad Ti\two/Sc\two\ absorption feature over the range
4000--4500\,\AA.

An impressive match is found for SN~2003gs, a low-luminosity \snia\ in
the optical bands (though slightly more luminous than 91bg-like
\sneia).  We assume the time of UVOIR maximum to
coincide with the first photometric measurement (which in turn
corresponds to the inferred time of $V$-band maximum by
\citealt{Krisciunas/etal:2009}) since there are no pre-maximum
observations for this SN. We expect the impact on the derived ratio of
peak UVOIR luminosities ($Q_{\rm uvoir}=1.12$) to be minor, given the
coincidence within 1\,d of the times of UVOIR and $V$-band maxima in
model DDC22. We checked that using the corrected $U$-band
photometry recently published by \cite{Krisciunas/etal:2012} had no
noticeable impact on the derived $Q_{\rm uvoir}$.

While the flux in the blue wing of Ca\two\,H\&K compares better than
for model DDC20 versus SN~2007hj, the predicted HVFs in the
Ca\two\ near-IR triplet are not present in the observations.

The spectroscopic features of SN~1998de \citep{Matheson/etal:2008} are
well matched by model DDC22, but the predicted luminosity is much too
high ($Q_{\rm uvoir}=2.5$ based on $BVRI$ photometry only).


\subsection{DDC25 ($M_{\rm \nifs}=0.18$\,\msun) vs. SN~1999by}

Model DDC25 corresponds to the classical low-luminosity 91bg-like
\sneia. The overall SED and individual spectral features match up
nicely with the well-observed 91bg-like SN~1999by
\citep{Garnavich/etal:2004}. We encounter the same problem as in
model DDC22 
concerning the prediction of prominent HVFs in the Ca\two\ near-IR
triplet.  The model appears slightly over-luminous with respect to
this SN ($Q_{\rm uvoir}=1.38$), but using only a $1\sigma$ larger
Cepheid distance to the host galaxy NGC~2841 (i.e. $\mu=30.97$\,mag
instead of $\mu=30.74$\,mag; \citealt{Macri/etal:2001}) results in a
better agreement, with $Q_{\rm uvoir}\approx 1.1$ (i.e. 10\% shift in
absolute flux). This model compares well to other 91bg-like \sneia,
such as the eponymous SN~1991bg
\citep{Filippenko/etal:1992b,Leibundgut/etal:1993} and SN~2005bl
\citep{Taubenberger/etal:2008}.

We thus confirm the result of \cite{Hoeflich/etal:2002}, that
low-luminosity \sneia\ can be explained within the framework of
Chandrasekhar-mass delayed-detonation models, contrary to claims otherwise
\citep[e.g.,][]{Pakmor/etal:2011}. We will need, however, to see how
our delayed-detonation models fare at pre-maximum phases through 
to the nebular phase.

\cite{Howell/etal:2001} show evidence for a significant intrinsic
polarization level of $\sim$0.3--0.8\% in SN~1999by near maximum
light, and argue that this SN has a well-defined symmetry axis,
possibly resulting from a rapidly-rotating single WD or a binary WD
merger process. Our excellent fits to the observed spectra with a 1D
radiative transfer code suggest that such asymmetries, if present, do not
dramatically affect the observables around peak bolometric luminosity.


\section{Discussion and conclusion}\label{sect:ccl}

In this paper, we have presented a first effort at modeling
\sneia\ with \cmfgen.  Our 1D approach solves the radiative-transfer
problem for SN~Ia ejecta with a high level of physical
consistency, which comes at a great computational expense.  The
coupled set of time-dependent statistical equilibrium, gas energy, and
radiative transfer (0$^{\rm th}$ and 1$^{\rm st}$ moments) equations
are solved for simultaneously, yielding multi-epoch SN~Ia spectra from
the UV to the infrared. The bolometric luminosity and the photometry are
obtained by suitable integrations of that wavelength-dependent
flux. Importantly, line blanketing is treated explicitly and
consistently --- the algorithm makes no assumption about the
thermalization character of the opacity as typically done in
approaches that do not solve for the level populations directly.

We compute radiative properties for a sequence of delayed-detonation
models \citep{Khokhlov:1991}, a mechanism in which a deflagration is
followed, after some delay depending on some prescribed transition
density, by a detonation in a Chandrasekhar-mass white dwarf. This
model is not new, and our investigation comes after nearly two decades
of study of this mechanism
\citep{Khokhlov:1991,Khokhlov/etal:1993,Hoeflich/Khokhlov/Wheeler:1995,
  Hoeflich/Khokhlov:1996,Hoeflich/etal:2002,Gamezo/Khokhlov/Oran:2005}.
However, we present results that have a higher level of physical
consistency than previously achieved, albeit in 1D.

Our models start at 1\,d after explosion and are evolved with
\cmfgen\ until beyond the bolometric peak. In this paper, we focus on
the phase of bolometric maximum of such delayed-detonation models.  We
analyze their properties and compare them with publicly-available
\snia\ data.  For each model in the delayed-detonation sequence, best
characterized by a \nifs\ mass ranging from 0.18 to 0.81\,\msun, we
generally find a \snia\ that fits the peak bolometric flux to within
$\sim$\,10\%, and the colors to within $\sim$0.1\,mag.  The spectral
features are typically reproduced with high fidelity, both in strength
and width.

Remarkably, with our set of delayed-detonation models, we reproduce
the basic morphological diversity of \snia\ radiative properties, from
blue, luminous, yet spectroscopically-normal \sneia\ (e.g.,
SN\,2009ig), to low-luminosity 91bg-like \sneia\ (e.g., SN\,1999by).
Our assessment is based primarily on optical data where the bulk of
the SN~Ia flux emerges, but we include UV data from the {\it Hubble
  Space Telescope} and the {\it Swift} satellite when available.  
We find that the rise time to bolometric maximum increases from
$\sim$\,15 to $\sim$21\,d with decreasing \nifs\ mass.  
Despite the complexity of our calculations, and the simplicity of
Arnett's rule \citep{Arnett:1979,Arnett:1982a}, we support its
validity to within 10\,\% in such delayed-detonation models.  Hence,
at its maximum, the bolometric luminosity of \sneia\ is close to the
instantaneous rate of decay energy, although both quantities are
directly related only at nebular phases (and physically equivalent in
the case of full $\gamma$-ray trapping).

We infer that the spectra of \sneia\ at maximum light primarily
reflect the change in ionization balance resulting from the magnitude
(and to a lesser extent, distribution) of heating from
\nifs\ decay. 
A reduction in temperature causes a reduction of the flux in the blue,
which is further exacerbated by a growing opacity from recombining
ions like Ti, Sc, or Fe.  This effect is particularly severe in the
$U$ and $B$ bands for models with a \nifs\ mass less than
$\sim$0.3\,\msun, and causes the transition to 91bg-like
\snia\ spectra.

Reproducing the line widths is more prone to error due to our 1D
approach, and the large ejecta kinetic energy in the explosion
models, which leads to broader lines than observed in the bulk of
\sneia. This problem seems however to affect mainly the Si\two\ lines, 
while others (e.g., Ca\two\,H\&K, which overlap with
Si\two\,3858\,\AA\ in the blue, and the S\two\ ``W'' feature around
5500\,\AA) match the observations. Across our model set, we reproduce
satisfactorily the widths of most features, the near-constant strength
of Si\two\,6355\,\AA, and the strong dependence of \rsi\ on
\nifs\ mass.

Ionization effects are particularly important for Ca\two\ lines. In
weaker explosions, Si\two\,6355\,\AA\ appears narrower when the
Ca\two\, near-IR triplet lines follow the opposite trend and appear
broader and stronger. While the former represents an interesting
diagnostic of the explosion energy, the latter is more sensitive to
ionization. In the delayed-detonation model with the least amount of
\nifs, ionization kinks give rise to high-velocity features in
Ca\,\two\ near-IR triplet: such ``structures'' are spherical shells of
distinct Ca ionization, producing optical-depth variations that affect
line profiles although they would produce no polarization signature.

In our model ejecta, there is essentially no carbon left by the
detonation, even if it is weak. This is not a shortcoming of the
delayed-detonation mechanism, but rather a problem with the assumption
of spherical symmetry for the explosion phase
\citep{Gamezo/Khokhlov/Oran:2005}. So, while ionization is key,
abundances can also completely inhibit the formation of certain lines.
The O\one\,7773\,\AA\ line is in principle a useful probe of
unburnt oxygen, although none of our maximum-light spectra shows
O\one\ absorption out to velocities where we recover the initial
$X({\rm O})\approx0.5$ composition of the progenitor WD. Nonetheless, our
models reproduce the observed trend of increasing
O\one\,7773\,\AA\ absorption with decreasing luminosity at maximum
light, and we obtain no O signature in our (1D) delayed-detonation
models that have a \nifs\ mass $\gtrsim0.4$\,\msun.

Even today, the inferences on the WD ejecta mass are not sufficiently
accurate to establish whether it is the Chandrasekhar mass, or more,
or less. At present, both $M_{\rm Ch}$ and sub-$M_{\rm Ch}$ models are
compatible with observations
\citep{Hoeflich/Khokhlov:1996,Sim/etal:2010}.  Hence, an answer to
this most fundamental question is still lacking. To identify a good
model for a SN~Ia, fitting the light curve is mandatory but it is
clearly not sufficient. Fitting the spectra is critical, and doing so
simultaneously, with the same physics, is an asset. However, one must
ensure a suitable match to observations at all times, not just at
maximum light as done in this paper. In particular, explosion models
are expected to best reveal their differences shortly after explosion
(e.g. through presence or absence of unburnt material) or at nebular
times when the IGE core (and possibly the innermost \nifs-deficient
region) is revealed.
  
Moreover, one is faced with the role of multi-D effects,
which may alter the spectral appearance sizeably for different viewing
angles \citep{KRW09}.  As we demonstrate here, assuming spherical
symmetry may still be suitable at certain epochs (e.g., bolometric
maximum).  However, 1D models miss important effects that are
inherently multi-dimensional, such as mixing. We can mimic the impact
of mixing in 1D by softening composition gradients, either
artificially \citep[e.g.,][]{Dessart/etal:2012a}, or based on angular
averages built from the few 3D simulations that exist, such as those
of \citet{Gamezo/Khokhlov/Oran:2005}. Performing multi-D
radiative-transfer simulations would seem warranted but all such
simulations so far have dramatically simplified the transfer
\citep[e.g.,][]{Kromer/Sim:2009}. One should question if the gain in
realism by going to 3D is not offset by the much simplified radiative
transfer compared to 1D.

An argument often made is that a given explosion model,
when post-processed by a radiative-transfer code, sometimes fails to
reproduce a given observation. One can then propose that this initial
explosion model is just not perfect. However, equally good
interpretations are that the radiative-transfer is missing some
important process, or that the atomic data is inaccurate. The
advantage of our strategy over such targeted studies is that if a
given mechanism is any good at reproducing \snia\ observations, we
should be able to find a number of perfect matches --- as we did. Out
of a few hundred \sneia, some must be pointing the ``right" way for
our model, or have a suitable chemical stratification, i.e., so that
what we adopt in our 1D model is a fair representation of what we see
averaged over all viewing angles. A viable model can fail for one SN
Ia, but should not fail for the few hundred \sneia\ for which we have
data today.

With the same set of delayed-detonation models, we will now
investigate the properties at earlier times (the brightening phase to
the peak of the light curve) and later times (transition past the peak
of the light curve into the nebular phase). Separately, we will
perform similar simulations for a wider range of progenitor and
explosion mechanisms, including pure deflagrations and pulsating
delayed detonations.


\section*{Acknowledgments}
LD acknowledges financial support from the European Community through
an International Re-integration Grant, under grant number
PIRG04-GA-2008-239184.  DJH acknowledges support from STScI theory
grants HST-AR-11756.01.A and HST-AR-12640.01 and NASA theory
grant NNX10AC80G. The work of AK was supported by the NSF
grant AST-0709181 (A.K.) ``Collaborative research: Three-Dimensional
Simulations of Type Ia Supernovae: Constraining Models 
with Observations.''  This work was granted access to the HPC
resources of CINES under the allocation 2011-c2011046608 and
2012-c2012046608 made by GENCI (Grand Equipement National de Calcul
Intensif). A subset of the computations were also performed at Caltech
Center for Advanced Computing Research on the cluster Zwicky funded
through NSF grant PHY-0960291 and the Sherman Fairchild
Foundation. 
The authors wish to thank: 
Eduardo Bravo for sending us his delayed-detonation model DDTc;
Ryan Chornock, Alex Filippenko, and Jeff Silverman, for sending us
spectra of SN~2007hj published by \cite{Silverman/etal:2012a}; 
Rubina Kotak, for sending us the spectrum of SN~2003gs used in this
paper ahead of publication; 
Ivo Seitenzahl, for providing neutronization rates in machine-readable
form.
This research has made use of the CfA
Supernova Archive, which is funded in part by the National Science
Foundation through grant AST 0907903; the Online Supernova Spectrum
Archive (SUSPECT); the Weizmann Interactive Supernova data
Repository (WISEREP, \citealt{WISEREP}).

\appendix

\section{Model Atoms}

Table~A1 gives the number of levels (both super-levels
and full levels; see \citealt{HM98} and \citealt{DH10} for details)
for the model atoms used in the radiative-transfer calculations
presented in this paper.

Oscillator strengths for CNO elements were originally taken from
\citet{NS83_LTDR, NS84_CNO_LTDR}. These authors also provide
transition probabilities to states in the ion continuum. The largest
source of oscillator data is from \citet{Kur09_ATD}\footnote{Data are
  available online at http://kurucz.harvard.edu}; its
principal advantage over many other sources (e.g., Opacity Project) is
that LS coupling is not assumed. More recently, non-LS oscillator
strengths have become available through the Iron Project
\citep{HBE93_IP}, and work done by the atomic-data group at Ohio State
University \citep{Nahar_OSU}. Other important sources of radiative
data for Fe include \citet{BB92_FeV, BB95_FeVI, BB95_FeIV},
\cite{Nahar95_FeII}.  Energy levels have generally been obtained from
National Institute of Standards and Technology.  Collisional data is
sparse, particularly for states far from the ground state. The
principal source for collisional data among low lying states for a
variety of species is the tabulation by \citet{Men83_col}; other
sources include \citet{BBD85_col}, \citet{LDH85_CII_col},
\citet{LB94_N2}, \citet{SL74}, \citet{T97_SII_col,T97_SIII_col}, Zhang
\& Pradhan (\citeyear{ZP95_FeII_col,ZP95_FeIII_col,ZP97_FeIV_col}).
Photoionization data is taken from the Opacity Project
\citep{Sea87_OP} and the Iron Project \citep{HBE93_IP}. Unfortunately
Ni and Co photoionization data is generally unavailable, and we have
utilized crude approximations. Charge exchange cross-sections are from
the tabulation by \citet{KF96_chg}.

\begin{table}
\begin{center}
\caption[]{Summary of the model atoms used in our radiative-transfer
  calculations. The source of the atomic data sets is given in
  \citet{DH10}.  N$_{\rm f}$ refers to the number of full levels,
  N$_{\rm s}$ to the number of super levels, and N$_{\rm trans}$ to
  the corresponding number of bound-bound transitions. The last column
  refers to the upper level for each ion treated. The total number of
  full (super) levels treated is 13\,959 (2149), which corresponds to
  629\,396 bound-bound transitions. Note that $n$w\,$^2$W refers to a
  state with principal quantum number $n$ (all $l$ states combined
  into a single state), and spin 2. Similarly, 8z\,$^1$Z refers to the
  $n=8$ state with high l states (usually $l=4$ and above) combined
  and spin 1.
\label{tab_atom_big}}
\begin{tabular}{l@{\hspace{3mm}}r@{\hspace{3mm}}r@{\hspace{3mm}}r@{\hspace{3mm}}l}
\hline
 Species        &  N$_{\rm f}$  &  N$_{\rm s}$ & N$_{\rm trans}$ & Upper Level \\
\hline
     C{\,\sc i}\,    &  26  &   14 &    120 & 2s2p$^3$$\,^3$P\opar\                  \\
     C{\,\sc ii}\,   &  26  &   14 &     87 & 2s$^2$4d\,$^2$D$_{5/2}$                \\
     C{\,\sc iii}\,  & 112  &   62 &    891 & 2s8f\,$^1$F\opar                       \\
     C{\,\sc iv}\,   &  64  &   59 &   1446 & $n=30$                         \\
     O{\,\sc i}\,    &  51  &   19 &    214 & 2s$^2$2p$^3$($^4$S\opar)4f\,$^3$F$_{3}$         \\
     O{\,\sc ii}\,   & 111  &   30 &   1157 & 2s$^2$2p$^2$($^3$P)4d\,$^2$D$_{5/2}$       \\
     O{\,\sc iii}\,  &  86  &   50 &    646 & 2p4f\,$^1$D                       \\
     O{\,\sc iv}\,   &  72  &   53 &    835 & 2p$^2$($^3$P)3p\,$^2$P\opar\                   \\
     Ne{\,\sc i}\,     & 139 &   70 &    1587 & 2s$^2$2p$^5$($^2$P\oparsub{3/2})6d\,$^2$[5/2]\oparsub{3}  \\
     Ne{\,\sc ii}\,  &  91  &   22 &   1106 & 2s$^2$2p$^4$($^3$P)4d\,$^2$P$_{3/2}$       \\
     Ne{\,\sc iii}\, &  71  &   23 &    460 & 2s$^2$2p$^3$($^2$D\opar)3d\,$^3$S\oparsub{1}         \\
     Na{\,\sc i}\,   &  71  &   22 &   1614 & 30w\,$^2$W                         \\
     Mg{\,\sc ii}\,  &  65  &   22 &   1452 & 30w\,$^2$W                         \\
     Mg{\,\sc iii}\, &  99  &   31 &    775 & 2p$^5$7s\,$^1$P\opar                   \\
     Al{\sc ii}\,  &  44  &   26 &    171 & 3s5d $^1$D$_{2}$                    \\
     Al{\sc iii}\, &  45  &   17 &    362 & 10z\,$^2$Z                         \\
     Si{\,\sc ii}\,  &  59  &   31 &    354 & 3s$^2$7g\,$^2$G$_{7/2}$             \\
     Si{\,\sc iii}\, &  61  &   33 &    310 & 3s5g\,$^1$G$_{4}$                    \\
     Si{\,\sc iv}\,  &  48  &   37 &    405 & 10f\,$^2$F\opar                       \\
     S{\,\sc ii}\,   & 324  &   56 &   8208 & 3s3p$^3$($^5$S\opar)4p\,$^6$P             \\
     S{\,\sc iii}\,  &  98  &   48 &    837 & 3s3p$^2$($^2$D)3d$\,^3$P             \\
     S{\,\sc iv}\,   &  67  &   27 &    396 & 3s3p($^3$P\opar)4p\,$^2$D$_{5/2}$         \\
     Ar{\,\sc i}      & 110 & 56  & 1541        & 3s$^2$3p$^5$($^2$P\oparsub{3/2})7p\,$^2$[3/2]$_2$  \\
     Ar{\,\sc ii}       & 415 & 134   & 20197  &     3s$^2$3p$^4$($^3$P$_1$)7i\,$^2$[6]$_{11/2}$  \\
     Ar{\,\sc iii}\, & 346  &   32 &   6898 & 3s$^2$3p$^3$($^2$D\opar)8s\,$^1$D\opar\            \\     
     Ca{\,\sc ii}\,  &  77  &   21 &   1736 & 3p$^6$30w\,$^2$W                     \\
     Ca{\,\sc iii}\, &  40  &   16 &    108 & 3s$^2$3p$^5$5s\,$^1$P\opar                \\
     Ca{\,\sc iv}\,  &  69  &   18 &    335 & 3s3p$^5$($^3$P\opar)3d\,$^4$D\oparsub{1/2}        \\
     Sc{\,\sc ii}\,  &  85  &   38 &    979 & 3p$^6$3d4f$\,^1$P\oparsub{1}             \\
     Sc{\,\sc iii}\, &  45  &   25 &    235 & 7h\,$^2$H\oparsub{11/2}                  \\
     Ti{\,\sc ii}\,  & 152  &   37 &   3134 & 3d$^2$($^3$F)5p\,$^4$D\oparsub{7/2}            \\
     Ti{\,\sc iii}\, & 206  &   33 &   4735 & 3d6f\,$^3$H\oparsub{6}                   \\
     Cr{\,\sc ii}\,  & 196  &   28 &   3629 & 3d$^4$($^3$G)4p\,x$^4$G\oparsub{11/2}          \\
     Cr{\,\sc iii}\, & 145  &   30 &   2359 & 3d$^3$($^2$D2)4p\,$^3$D\oparsub{3}             \\
     Cr{\,\sc iv}\,  & 234  &   29 &   6354 & 3d$^2$($^3$P)5p\,$^4$P\oparsub{5/2}            \\
     Mn{\,\sc ii}\,  &  97  &   25 &    236 & 3d$^4$($^5$D)4s$^2$\,c$^5$D$_{4}$            \\
     Mn{\,\sc iii}\, & 175  &   30 &   3173 & 3d$^4$($^3$G)4p\,y$^4$H\oparsub{13/2}         \\
     Fe{\,\sc i}\,   & 136  &   44 &   1900 & 3d$^6$($^5$D)4s4p\,$^5$F\oparsub{3}           \\
     Fe{\,\sc ii}\,  & 827  &  275 &  44\,831 & 3d$^5$($^6$S)4p$^2$($^3$P)\,$^4$P$_{1/2}$        \\
     Fe{\,\sc iii}\, & 607  &   69 &   9794 & 3d$^5$($^4$D)6s$^3$D$_{2}$               \\
     Fe{\,\sc iv}\,  &1000  &  100 &  72\,223 & 3d$^4$($^3$G)4f\,$^4$P\oparsub{5/2}            \\
     Fe{\,\sc v}\,   & 191  &   47 &   3977 & 3d$^3$($^4$F)4d\,$^5$F$_{3}$              \\
     Fe{\,\sc vi}\,  & 433  &   44 &  14\,103 & 3p$^5$($^2$P\opar)3d$^4$($^1$S)\,$^2$P\oparsub{3/2}      \\
     Co{\,\sc ii}\,  &1000  &   81 &  61\,986 & 3d$^7$($^4$P)4f\,$^5$F\oparsub{4}             \\
     Co{\,\sc iii}\, &1000  &   72 &  68\,462 & 3d$^6$($^5$D)5f\,$^4$F\oparsub{9/2}            \\
     Co{\,\sc iv}\,  &1000  &   56 &  69\,425 & 3d$^5$($^2$D)5s\,$^1$D$_{2}$              \\
     Co{\,\sc v}\,   & 387  &   32 &  13\,605 & 3d$^4$($^3$F)4d\,$^2$H$_{9/2}$            \\
    Co{\,\sc vi}\,  & 323  &   28 &   9608 & 3d$^3$($^2$D)4d$\,^1$S$_{0}$ \\
     Ni{\,\sc ii}\,  &1000  &   59 &  51\,707 & 3d$^8$($^3$F)7f\,$^4$I\oparsub{9/2}            \\
     Ni{\,\sc iii}\, &1000  &   47 &  66\,486 & 3d$^7$($^2$D)4d\,$^3$S$_{1}$b             \\
     Ni{\,\sc iv}\,  &1000  &   54 &  72\,898 & 3d$^6$($^5$D)6p\,$^6$F\oparsub{11/2}           \\
     Ni{\,\sc v}\,   & 183  &   46 &   3065 & 3d$^5$($^2$D3)4p\,$^3$F\oparsub{3}            \\
     Ni{\,\sc vi}\,  & 314  &   37 &   9569 & 3d$^4$($^5$D)4d\,$^4$F$_{9/2}$            \\
\hline
\end{tabular}
\end{center}
\end{table}

\section{Contribution of individual ions to the total flux}\label{sect:ladder}

In what follows we illustrate the contribution of individual ions
to the full synthetic spectra, compared to the same observed \snia\
spectra as in Fig.~\ref{fig:comp_maxspec}. The upper pannel in each
plot shows the full model (black) and the observed spectrum
(red). Labels are the same as in Fig.~5.
The lower pannel in each plot shows the contribution of individual
ions, computed by taking the ratio of the full spectrum
($F_{\lambda,{\rm All}}$) to that excluding all bound-bound
transitions of the corresponding ion ($F_{\lambda,{\rm less}}$). 

We selected ions that contribute significantly to the observed total
flux in at least one of the delayed-detonation models at bolometric
maximum. The following ions showed 
essentially no noticeable contribution in the wavelength range
considered here (2000--10000\,\AA): C\one--{\sc ii}, O\two, Ne\two,
Na\one, Mg\one, Al\two--{\sc iii}, Ar\three, Sc\three, Ti\three,
Cr\three, Mn\two, and Ni\three.

These plots are mostly for educational purposes, yet reveal key
observed trends with decreasing \nifs\ mass: increase in Ca\two\
strength (with double-absorption features resulting from 
optical-depth effects alone; see Sect.~5.4);
increase of \rsi\ ratio and decrease of 
Si\two\,6355\,\AA\ absorption velocity and width; sudden onset of
prominent Ti\two/Sc\two\ blanketing in the 4000--4500\,\AA\
region at the transition to low-luminosity \sneia; increase in
strength of the O\one\,7773\,\AA\ line.

\begin{figure*}
\centering
\epsfig{file=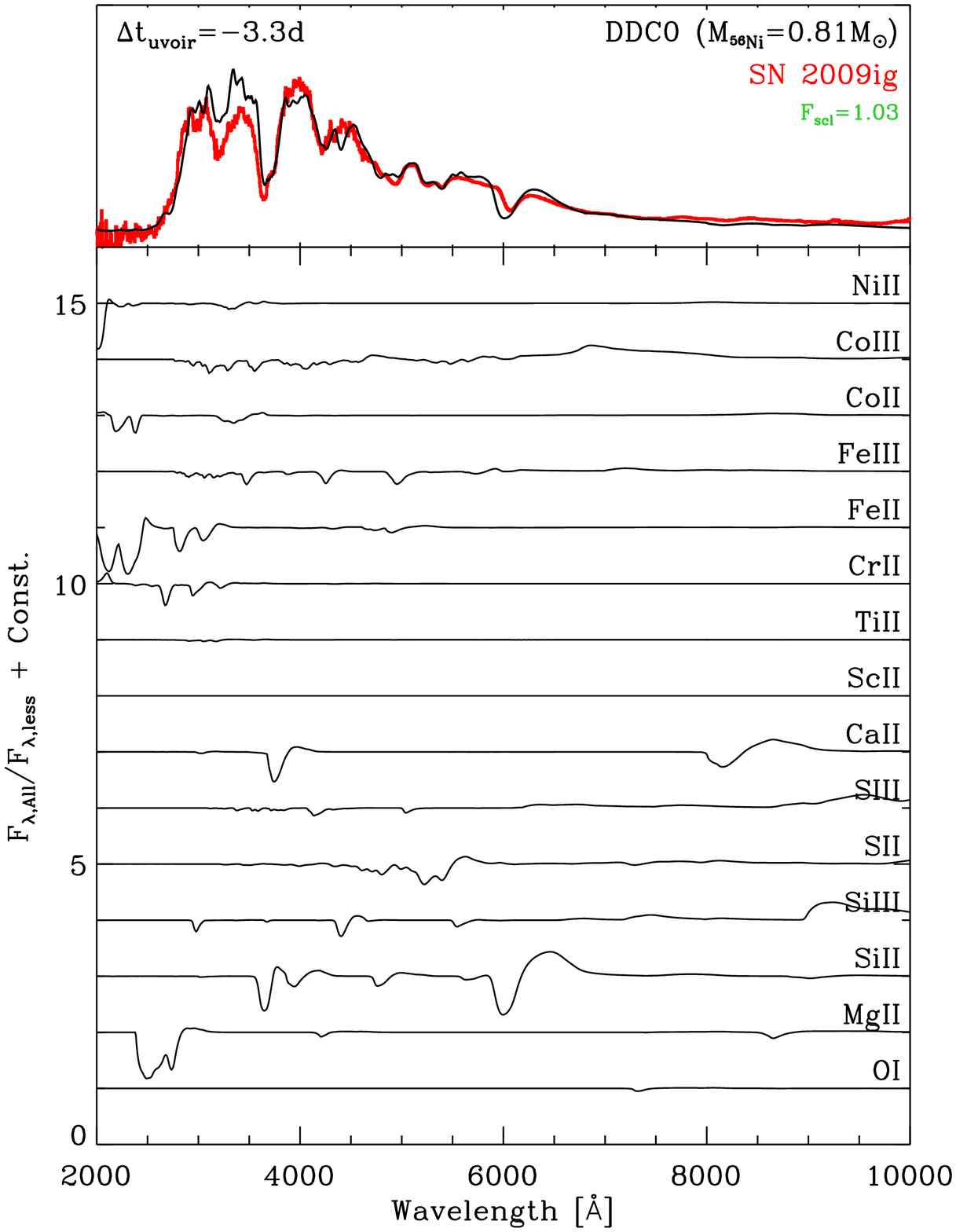,width=.97\textwidth}
\caption{\label{fig:comp_maxspec_DDC0_ladder}Contribution of
    individual ions (bottom) to the full synthetic spectrum of DDC0
    (top, black line), compared to SN~2009ig at $-3.3$\,d from UVOIR
    maximum (top, red line).}
\end{figure*}

\begin{figure*}
\centering
\epsfig{file=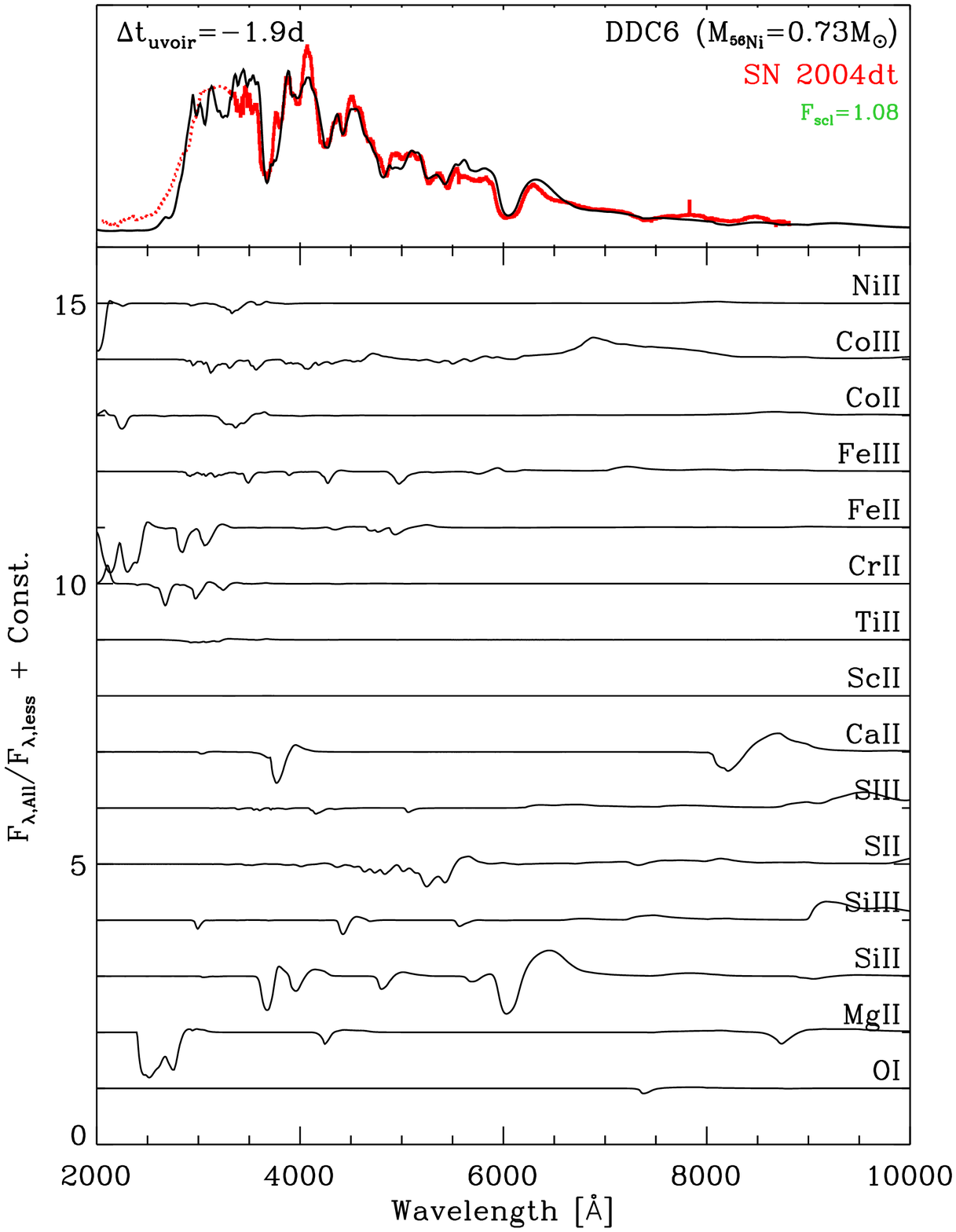,width=.97\textwidth}
\caption{\label{fig:comp_maxspec_DDC6_ladder}Contribution of
  individual ions (bottom) to the full synthetic spectrum of DDC6
  (top, black line), compared to SN~2004dt at $-1.9$\,d from UVOIR maximum
  (top, red line).}
\end{figure*}

\begin{figure*}
\centering
\epsfig{file=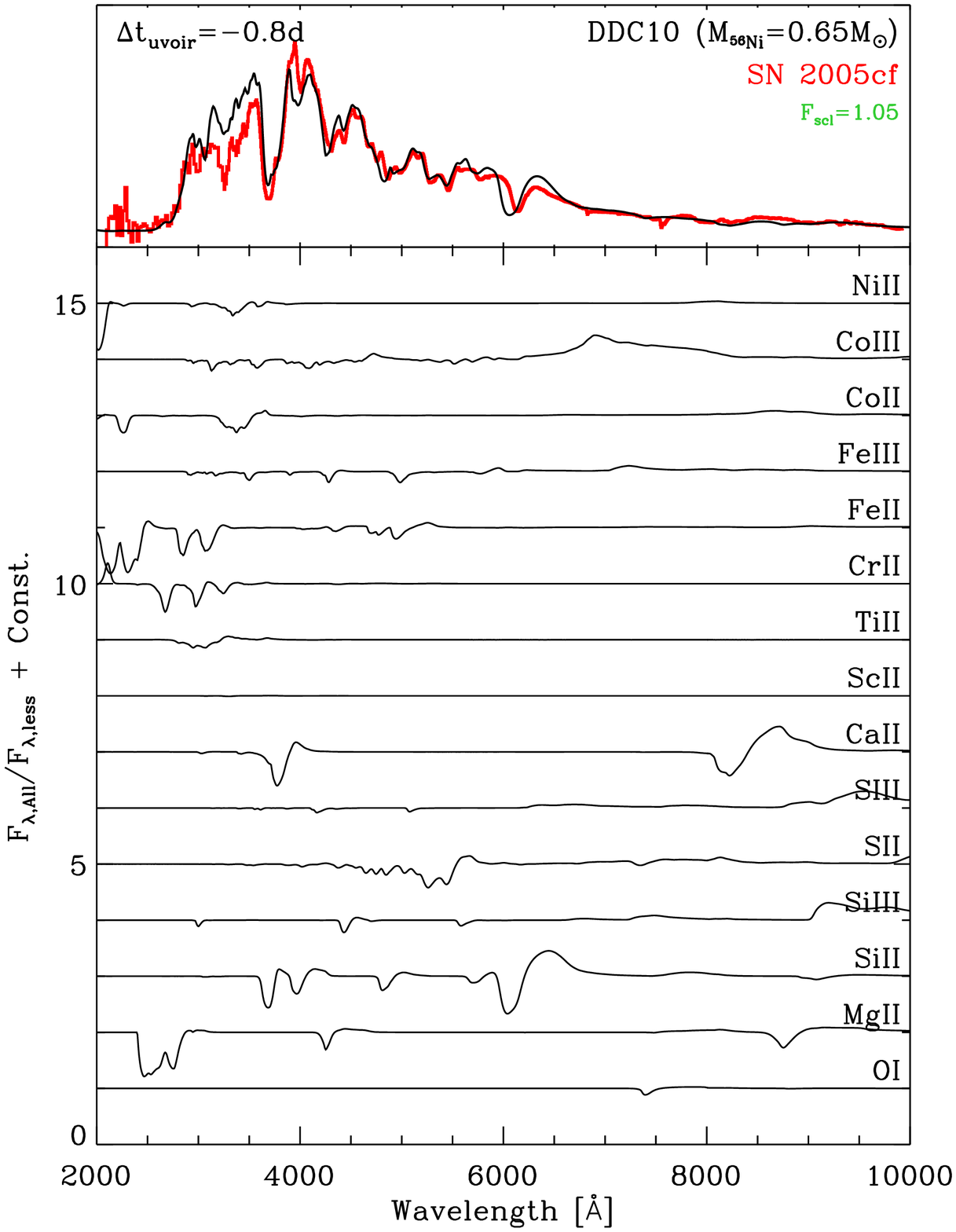,width=.97\textwidth}
\caption{\label{fig:comp_maxspec_DDC10_ladder}Contribution of
  individual ions (bottom) to the full synthetic spectrum of DDC10
  (top, black line), compared to SN~2005cf at $-0.8$\,d from UVOIR maximum
  (top, red line).}
\end{figure*}

\begin{figure*}
\centering
\epsfig{file=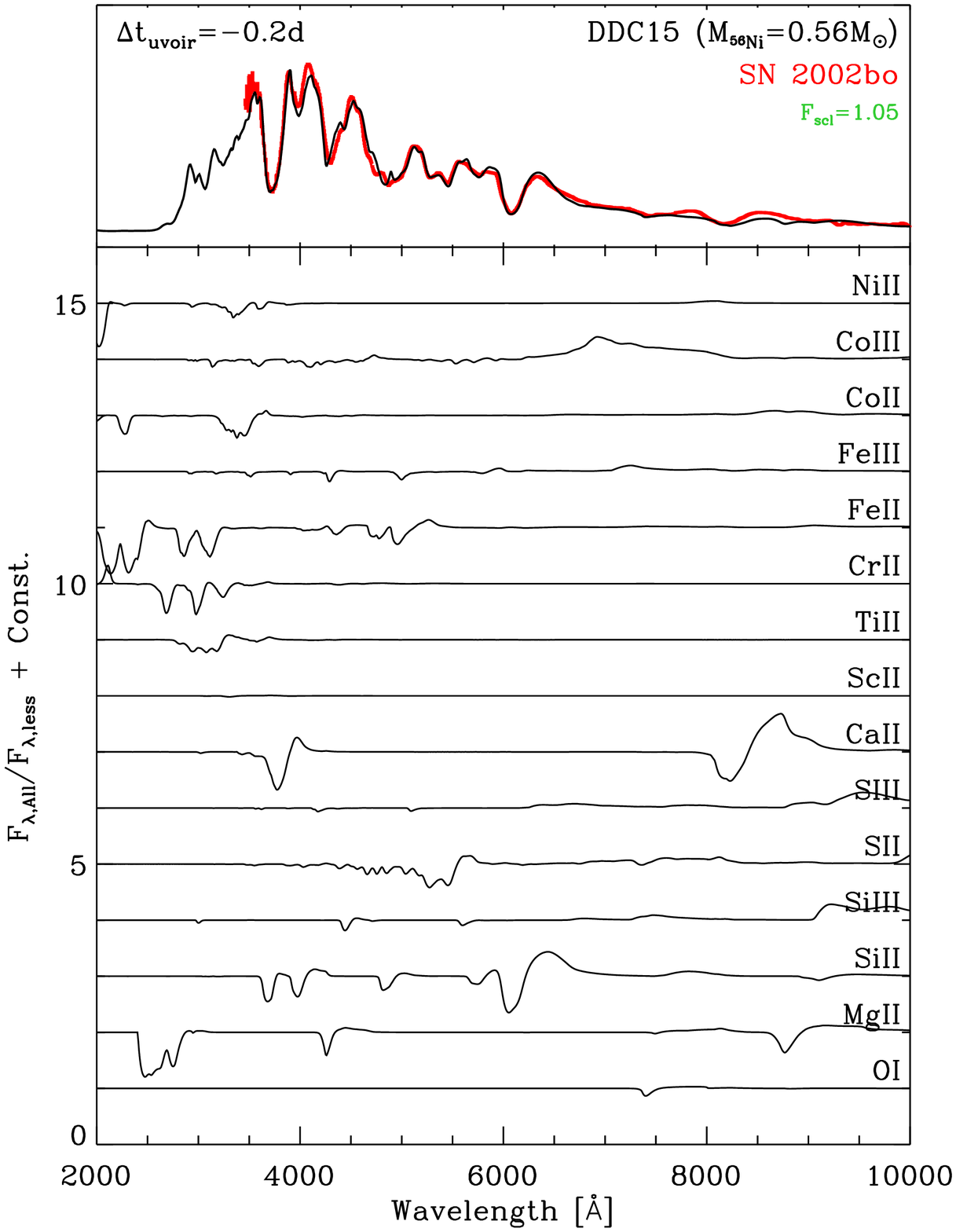,width=.97\textwidth}
\caption{\label{fig:comp_maxspec_DDC15_ladder}Contribution of
  individual ions (bottom) to the full synthetic spectrum of DDC15
  (top, black line), compared to SN~2002bo at $-0.2$\,d from UVOIR maximum
  (top, red line).}
\end{figure*}

\begin{figure*}
\centering
\epsfig{file=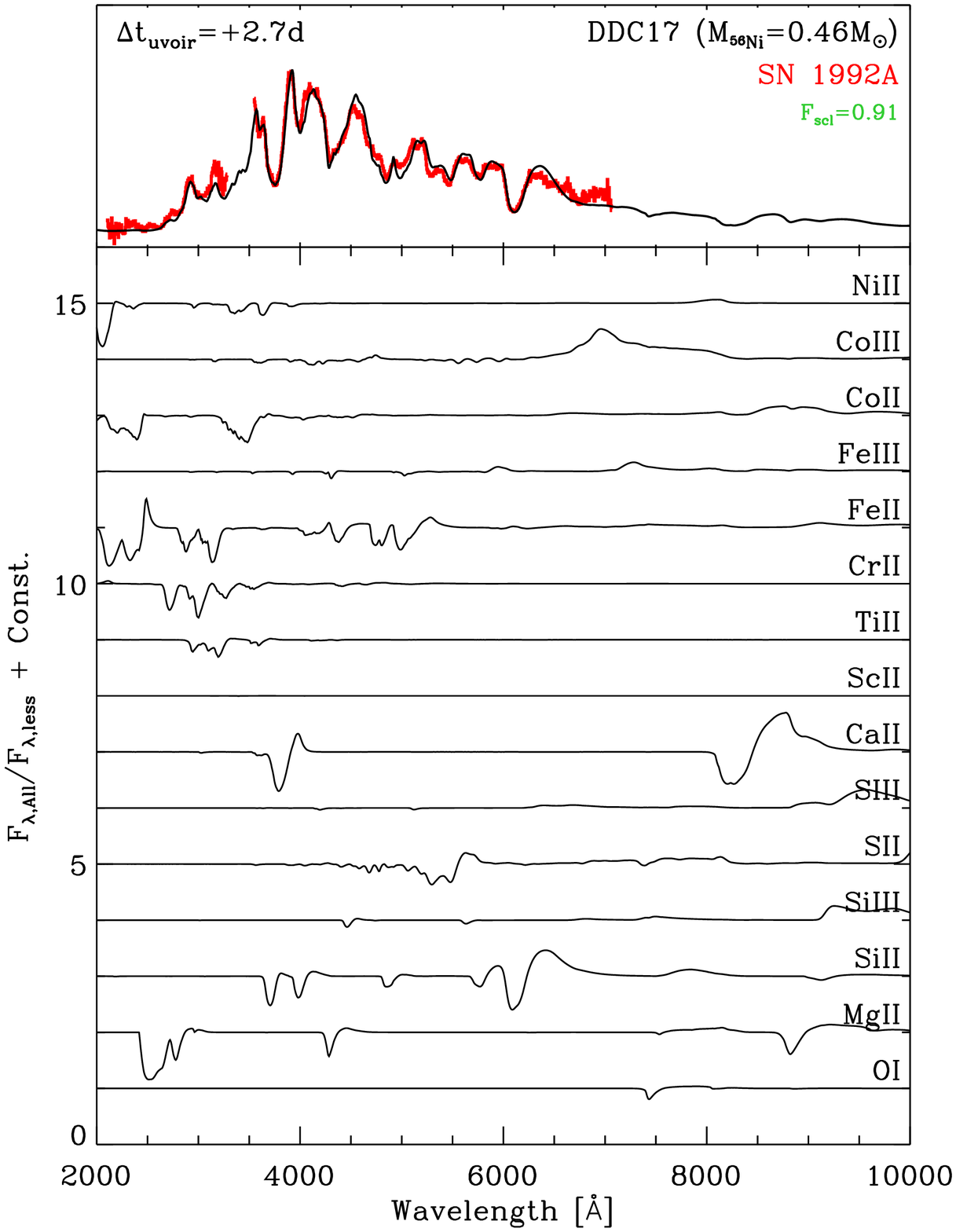,width=.97\textwidth}
\caption{\label{fig:comp_maxspec_DDC17_ladder}Contribution of
  individual ions (bottom) to the full synthetic spectrum of DDC17
  (top, black line), compared to SN~1992A at +2.7\,d from UVOIR maximum
  (top, red line).}
\end{figure*}

\begin{figure*}
\centering
\epsfig{file=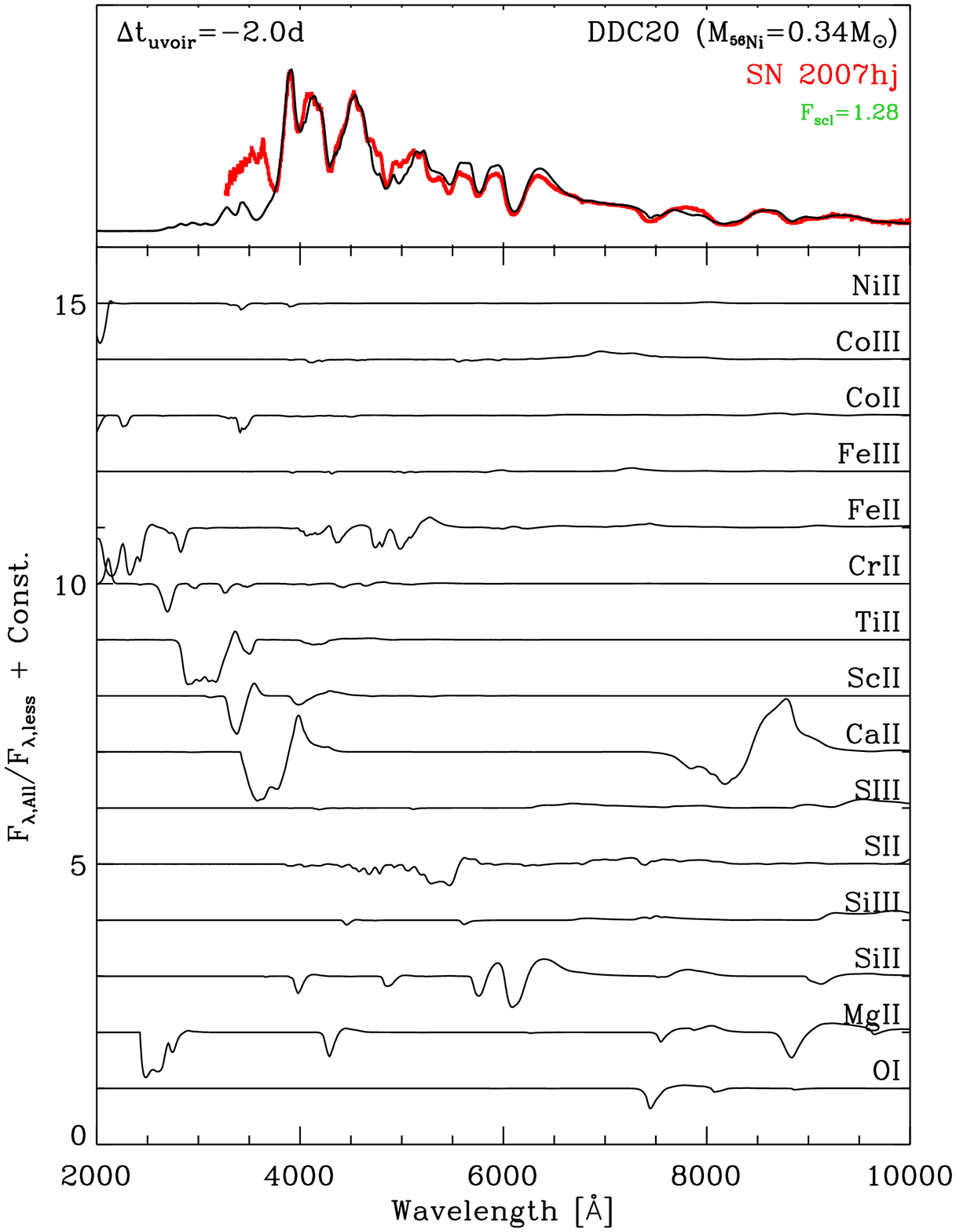,width=.97\textwidth}
\caption{\label{fig:comp_maxspec_DDC20_ladder}Contribution of
  individual ions (bottom) to the full synthetic spectrum of DDC20
  (top, black line), compared to SN~2007hj at $-2.0$\,d from UVOIR maximum
  (top, red line).}
\end{figure*}

\begin{figure*}
\centering
\epsfig{file=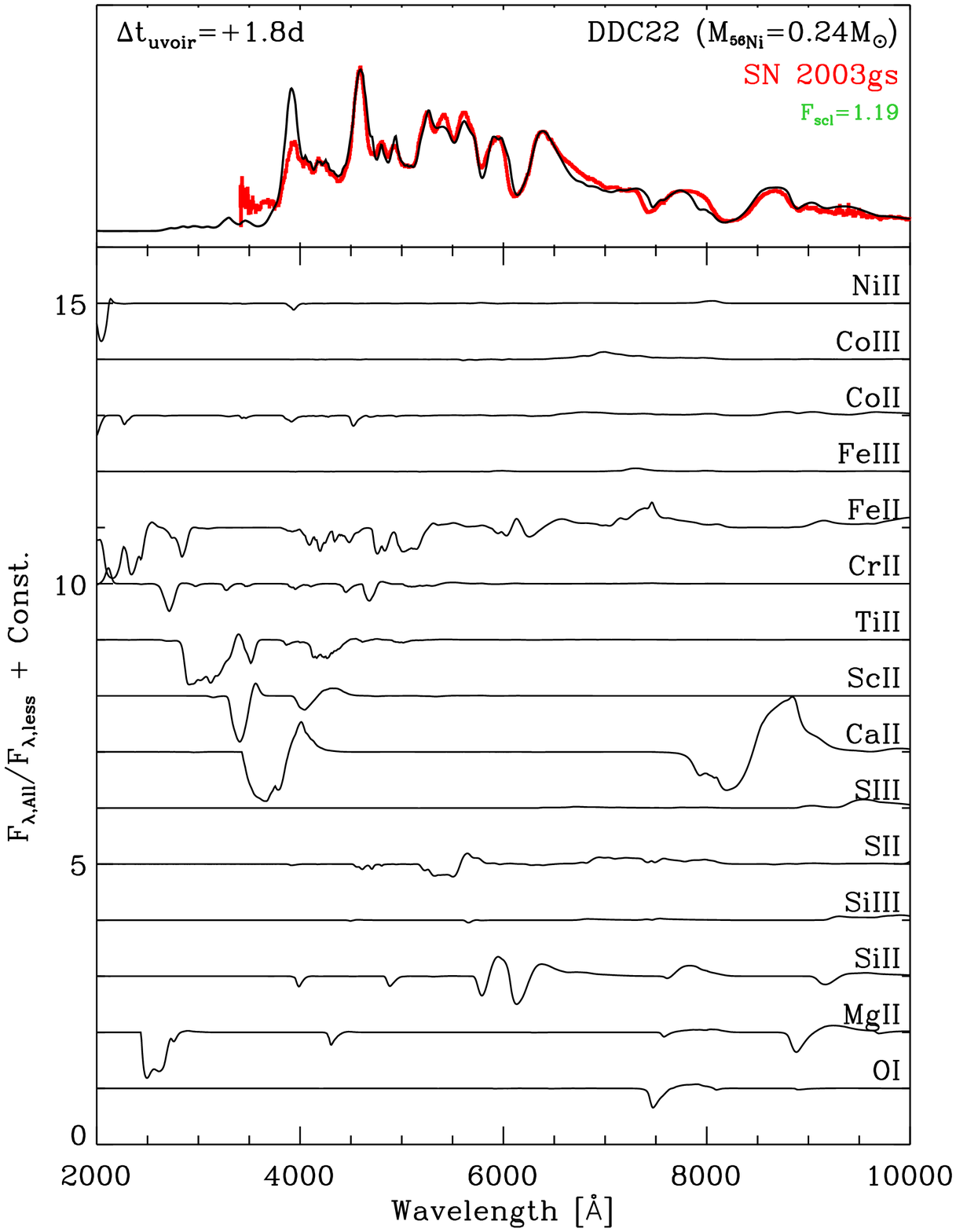,width=.97\textwidth}
\caption{\label{fig:comp_maxspec_DDC22_ladder}Contribution of
  individual ions (bottom) to the full synthetic spectrum of DDC22
  (top, black line), compared to SN~2003gs at +1.8\,d from UVOIR maximum
  (top, red line).}
\end{figure*}

\begin{figure*}
\centering
\epsfig{file=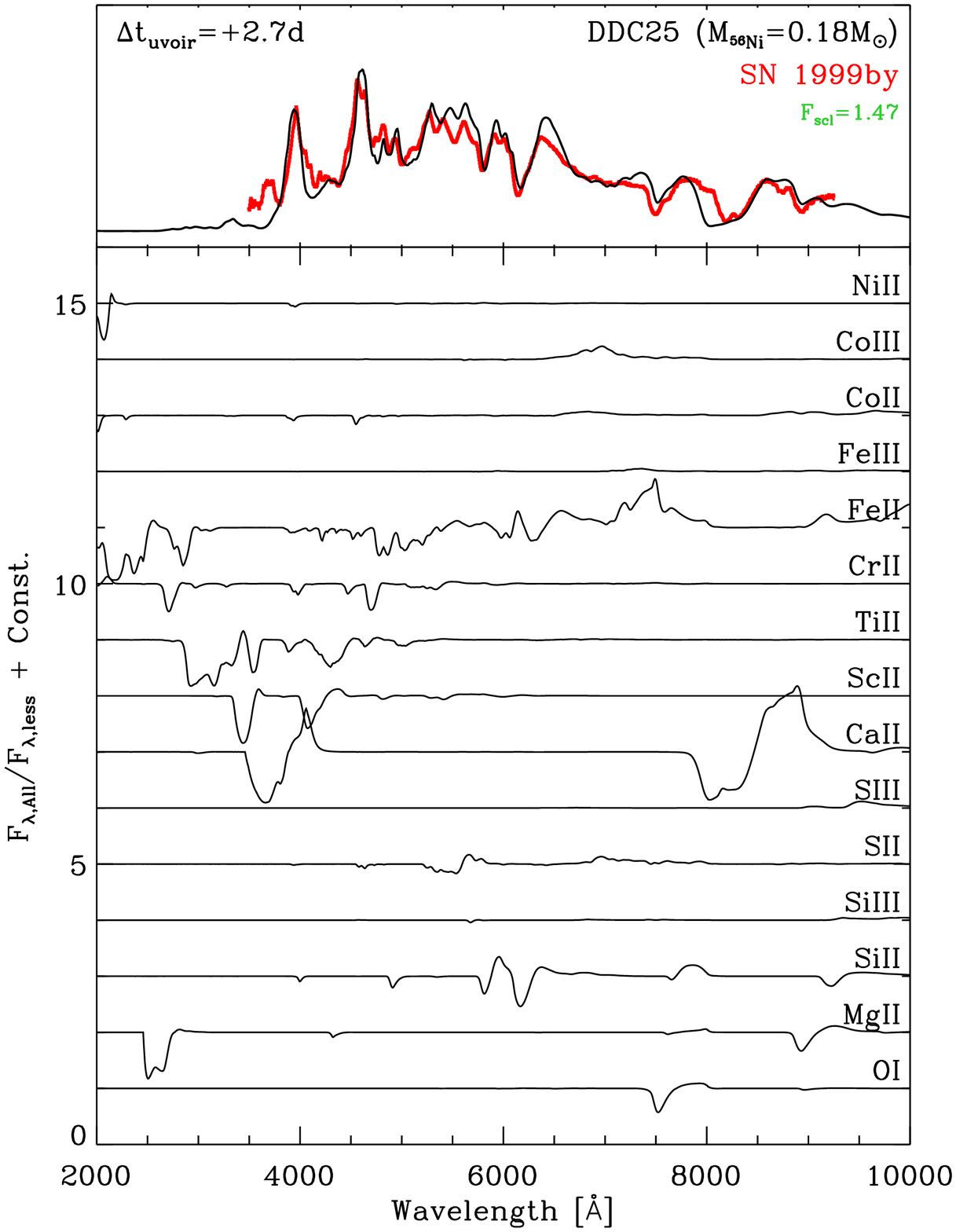,width=.97\textwidth}
\caption{\label{fig:comp_maxspec_DDC25_ladder}Contribution of
  individual ions (bottom) to the full synthetic spectrum of DDC25
  (top, black line), compared to SN~1999by at +2.7\,d from UVOIR maximum
  (top, red line).}
\end{figure*}

\section{Log of bolometric luminosity and absolute magnitudes at
  bolometric maximum}

Table~C1 lists the bolometric luminosity and
$UBVRIJHK$ magnitudes of our models at bolometric maximum.
The last line gives their standard deviation amongst our model
series. Note the small scatter in the NIR bands with respect to the
optical.

\begin{table*}
\caption{Bolometric luminosity and absolute $UBVRIJHK$ magnitudes at bolometric maximum in our model set.}\label{tab:lbolpeakmags}
\begin{tabular}{lccccccccc}
\hline\hline
\multicolumn{1}{c}{Model} & $L_{\rm bol}$  & $M_U$ & $M_B$ & $M_V$ & $M_R$ & $M_I$ & $M_J$ & $M_H$ & $M_K$ \\
                          & [erg s$^{-1}$] & [mag] & [mag] & [mag] & [mag] & [mag] & [mag] & [mag] & [mag] \\
\hline
DDC0            & 1.77(43) & $-$20.16 & $-$19.53 & $-$19.51 & $-$19.36 & $-$18.67 & $-$18.14 & $-$17.41 & $-$17.56 \\
DDC6            & 1.62(43) & $-$20.05 & $-$19.44 & $-$19.41 & $-$19.31 & $-$18.66 & $-$18.18 & $-$17.44 & $-$17.52 \\
DDC10           & 1.45(43) & $-$19.91 & $-$19.35 & $-$19.30 & $-$19.26 & $-$18.69 & $-$18.26 & $-$17.54 & $-$17.56 \\
DDC15           & 1.22(43) & $-$19.63 & $-$19.20 & $-$19.16 & $-$19.18 & $-$18.73 & $-$18.33 & $-$17.68 & $-$17.64 \\
DDC17           & 1.03(43) & $-$19.28 & $-$19.03 & $-$19.03 & $-$19.09 & $-$18.70 & $-$18.33 & $-$17.78 & $-$17.67 \\
DDC20           & 7.65(42) & $-$18.42 & $-$18.75 & $-$18.84 & $-$18.91 & $-$18.61 & $-$18.34 & $-$17.99 & $-$17.78 \\
DDC22           & 5.10(42) & $-$17.34 & $-$17.99 & $-$18.51 & $-$18.64 & $-$18.47 & $-$18.28 & $-$18.17 & $-$17.90 \\
DDC25           & 3.64(42) & $-$16.65 & $-$17.40 & $-$18.19 & $-$18.36 & $-$18.20 & $-$18.07 & $-$18.04 & $-$17.72 \\
\hline
StdDev          & 5.16(42) & \multicolumn{1}{r}{    1.33} & \multicolumn{1}{r}{    0.76} & \multicolumn{1}{r}{    0.46} & \multicolumn{1}{r}{    0.35} & \multicolumn{1}{r}{    0.18} & \multicolumn{1}{r}{    0.10} & \multicolumn{1}{r}{    0.29} & \multicolumn{1}{r}{    0.13} \\
\hline
\end{tabular}

\flushleft
{\bf Note:} Numbers in parentheses correspond to powers of ten.
\end{table*}


\bibliographystyle{mn2e}
\bibliography{ms_maxspec,atomic}

\label{lastpage}

\end{document}